\documentclass[sigconf]{acmart}

\AtBeginDocument{%
  \providecommand\BibTeX{{%
    \normalfont B\kern-0.5em{\scshape i\kern-0.25em b}\kern-0.8em\TeX}}}

\setcopyright{acmcopyright}
\copyrightyear{2018}
\acmYear{2018}
\acmDOI{10.1145/1122445.1122456}

\usepackage{ifthen}
\usepackage{color}
\usepackage{comment}
\usepackage{svg}
\usepackage[]{algorithm2e}
\usepackage{tabularx}

\newboolean{showcomments}
\setboolean{showcomments}{true} 
\ifthenelse{\boolean{showcomments}}
  {\newcommand{\nb}[2]{
    \fcolorbox{gray}{yellow}{\bfseries\sffamily\scriptsize#1}
    {\sf\small$\blacktriangleright$\textit{#2}$\blacktriangleleft$}
   }
   
  }
  {\newcommand{\nb}[2]{}
   
  }

\newcommand{\figpng}[3]{ 
    \begin{figure}[tb]
        \centerline{\includegraphics[width=#2\textwidth]{#1.png}}
        \caption{#3}
        \label{fig:#1}
    \end{figure} 
}
\newcommand{\figpdf}[3]{ 
    \begin{figure}[tb]
        \centerline{\includegraphics[width=#2\textwidth]{#1.pdf}}
        \caption{#3}
        \label{fig:#1}
    \end{figure} 
}

\newcommand{\mar}[0] {{\small\sffamily\textsc{MAR}}\xspace}

\newcommand{\code}[1] {{\small\sffamily #1}}

\usepackage{array}
\newcolumntype{H}{>{\setbox0=\hbox\bgroup}c<{\egroup}@{}}

\usepackage{tikz}
\usepackage{calc}

\usetikzlibrary{arrows}
\usetikzlibrary{shapes}

\newcommand{\cpath}[1]{%
  {\footnotesize\sffamily $\Big(\,${#1}$\,\Big)$}
  }

\newcommand{\att}[1]{%
  \tikz[baseline=(char.base)]\node[anchor=south west, draw,ellipse, inner sep=0.5pt, minimum size=4mm,
    text height=2mm](char){{#1}} ;}

\newcommand{\cl}[1]{%
  \tikz[baseline=(char.base)]\node[anchor=south west, draw,rectangle, rounded corners, inner sep=2pt, minimum size=4mm,
    text height=2mm](char){{#1}} ;}

\newcommand{\edge}[1]{%
  $\overrightarrow{\text{{#1}}}$}

\usepackage{pgf}
\usepackage{pgfmath}

\usetikzlibrary{calc}
\usepackage{amsmath,array,booktabs}

\copyrightyear{2020}
\acmYear{2020}
\setcopyright{acmlicensed}\acmConference[MODELS '20]{ACM/IEEE 23rd International Conference on Model Driven Engineering Languages and Systems}{October 18--23, 2020}{Montreal, QC, Canada}
\acmBooktitle{ACM/IEEE 23rd International Conference on Model Driven Engineering Languages and Systems (MODELS '20), October 18--23, 2020, Montreal, QC, Canada}
\acmPrice{15.00}
\acmDOI{10.1145/3365438.3410947}
\acmISBN{978-1-4503-7019-6/20/10}

\begin{document}



\title{MAR: A structure-based search engine for models}



\author{Jos\'e Antonio Hern\'andez L\'opez}
\affiliation{%
  \institution{Universidad de Murcia}
}
\email{joseantonio.hernandez6@um.es}

\author{Jes\'us S\'anchez Cuadrado}
\affiliation{\institution{Universidad de Murcia}}
\email{jesusc@um.es}




\begin{abstract}
The availability of shared software
models provides opportunities for reusing, adapting and learning from
them. Public models are typically stored in a variety of locations, including
model repositories, regular source code repositories, web pages, etc. 
To profit from them developers need effective search mechanisms to locate the models relevant
for their tasks. However, to date, there has been little success in creating a generic and efficient search engine specially tailored to the modelling domain. 

In this paper we present MAR, a search engine for models. MAR is generic 
in the sense that it can index any type of model if its meta-model is known. MAR uses a query-by-example approach, that is, it uses example
models as queries. The search takes the model structure into account using the notion of bag of paths, which encodes the structure of a model using paths between model elements and is a representation amenable for indexing.
MAR is built over HBase using a specific design to deal with large repositories.
Our benchmarks
show that the engine is efficient and has fast response times
in most cases.
We have also evaluated the precision of the search engine by creating
model mutants which simulate user queries. 
A REST API is available to perform queries and an Eclipse plug-in allows end users to connect
to the search engine from model editors. 
We have currently indexed more than 50.000 models of different kinds, including
Ecore meta-models, BPMN diagrams and UML models. MAR is available at~\url{http://mar-search.org}.






\end{abstract}



\maketitle

\section{Introduction}



Model repositories are essential for the success
of the MDE paradigm~\cite{bucchiarone2020grand, di2015collaborative}, since they have the potential to
foster communities of modellers, improve learning by letting
newcomers explore existing models and provide high-quality models which can be reused.
There are several model repositories available~\cite{di2015collaborative},
some of which offer public models. 
For instance, the GenMyModel\footnote{\url{https://www.genmymodel.com/}} cloud modeling service features a public
repository with thousands of available models. To profit from them
advanced search mechanisms are needed~\cite{basciani2015model}.
In particular, a search engine has the potential to 
aggregate models from several sources, including 
model repositories (e.g., the AtlanMod Zoo\footnote{\url{https://web.imt-atlantique.fr/x-info/atlanmod/index.php?title=Zoos}}) and code repositories (e.g., GitHub~\cite{robles2017extensive}), thus
boosting the user chances to retrieve the desired results easily.
Despite some efforts~\cite{bislimovska2014textual, lucredio2012moogle, gomes2004using} there are still open challenges,
which includes indexing models efficiently for search and retrieval, taking the model structure into account in the search, and
aggregating models of different types and from different sources under a single plataform.



In this paper we present \mar, a search engine specifically designed for models. \mar is generic, in the sense that it can handle any EMF model~\cite{steinberg2008emf}
regardless of its meta-model. An important challenge is how
to encode the structure of the model in a manner that can
be stored in an inverted index~\cite{arasu2001searching} (a map from items to documents containing such item). To tackle this we encode
a model by computing paths of 
maximum, configurable length $n$ between objects and attribute values
of the model.
At the user level, our system employs a query-by-example approach in which the user
just needs to create model fragments using a regular modelling tool. The system
extracts the paths of the query and access the inverted index to perform the
scoring of the relevant models.
We have evaluated the precision of \mar
for finding relevant models by automatically deriving queries using mutation operators,
obtaining good results. Moreover, we evaluate its performance  obtaining results that show its practical applicability.
We also discuss several applications of \mar beyond its main use case, that is, searching for models stored in repositories. Notably
we have used MAR for meta-model classification
showing that using a simple $k-$Nearest Neighbours approach can reach similar precision as state-of-the-art methods~\cite{nguyen2019automated}.
Finally, \mar has currently indexed more than 50.000 models
from several sources, including Ecore meta-models, UML models and BPMN models. \mar is freely accessible at \url{http://mar-search.org}
through a REST API, a web interface and an Eclipse plug-in.




%


\vspace{2pt}
{\noindent \bf Organization.} 
Section~\ref{sec:overview} presents an overview of our approach and the
running example. The formal model of the search process is presented in Sect.~\ref{sec:model_retrieving}, whereas Sect.~\ref{sec:components} describes the practical aspects in the design of our search engine. Section~\ref{sec:applications} introduces applications of a search engine, while Sect.~\ref{sec:evaluation} reports the results of the
evaluation. Finally, Sect.~\ref{sec:related} discusses the related work and Sect.~\ref{sec:conclusions} concludes.

\section{Overview}\label{sec:overview}


The design of a search engine for models requires considering several aspects~\cite{lucredio2012moogle}. The search should be {\em meta-model based} in the sense that only models conforming to the meta-model of interest are retrieved. 
There exists search engines for specific modelling languages (e.g., UML~\cite{gomes2004using} and WebML~\cite{bislimovska2014textual}), but the diversity of modelling approaches and the emergence of DSLs suggests that search engines should be generic.
Moreover, the nature of the search process requires an algorithm to perform {\em inexact matching} because the user is interested in obtaining several search results, using
some {\em ranking mechanism} to sort the results according to their relevance.
A search engine requires an {\em indexing} mechanism for processing and storing the available models in a manner which is adequate for efficient look up.
In addition, a good search engine should be able to handle {\em large repositories} of models
while maintaining a good performance as well as search precision.
Another aspect is how to present the results to the user. This requires considering the
integration with existing tools and building dedicated web services.


\figpdf{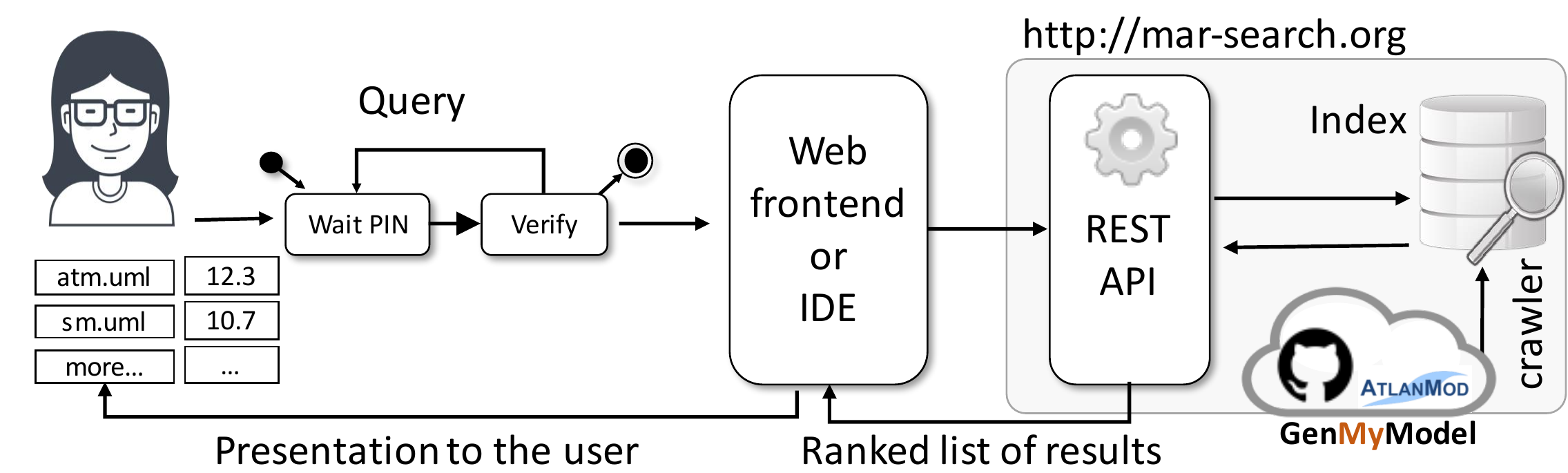}{0.5}{Usage and components of the search engine.}

Our design addresses these concerns by including the following features, which are depicted in Fig.~\ref{fig:overview-usage}.
\begin{itemize}
    \item {\bf Query by example}. The user interacts with the system using a query mechanism  based on examples. 
    The user creates an example model containing elements that describe
    the expected results.
    The system receives the model (and the name of its meta-model) to drive the search and to present a ranked list of results.
    
    \item {\bf Frontend}. The functionality is exposed through a REST API which can be used directly
    or through some tool exposing its capabilities in a user friendly manner. In particular, we have implemented a web interface and an Eclipse plug-in which integrates the
    search engine with Ecore editors.

    
    \item {\bf Indexing}. An inverted index is populated with models crawled from existing software
          repositories and datasets (e.g., Github, GenMyModel). Implementation-wise, it is an HBase database,
          which stores the information needed for the search. 

    \item {\bf Scoring algorithm}. The results are obtained by accessing the index to collect 
          the relevant models among those available. They are ranked according to a scoring function and returned to the
          user. The IDE or the web frontend are responsible for presenting the results in
          a user-friendly manner.

    \item {\bf Generic search}. The search is meta-model based because it takes into account the structure given by the meta-model, but the engine is generic, in the sense that it can index and search models conforming to any EMF meta-model. 

\end{itemize}



{\noindent {\bf Running example.}} 
Let us consider the task of searching
for UML state machine models using a search engine, like \mar, which has indexed thousands of UML models.
An excerpt of the UML meta-model for state machine diagrams is shown in Fig.~\ref{fig:statemachine-metamodel}.
Let us suppose that \mar has indexed a state machine which represents making or receiving mobile phone calls, 
similar to the diagram shown in Fig.~\ref{fig:example_models}(left). 
A user interested in models representing phone calls may create a query like the one in Fig.~\ref{fig:example_models}(right).
The query is different since it models only the reception of calls. The modelling style is also different because in the query the system waits for the incoming call in \code{Wait} and then transitions to \code{Waiting to pick up}, whereas in Fig.~\ref{fig:example_models}(left)
the system automatically transitions to \code{Talking} when there is an incoming call.
A search engine for models should be able to match attribute values (e.g., strings) and to detect which structures of the model match to which structures of the query,
and provide a relevance score based on this. 

\figpdf{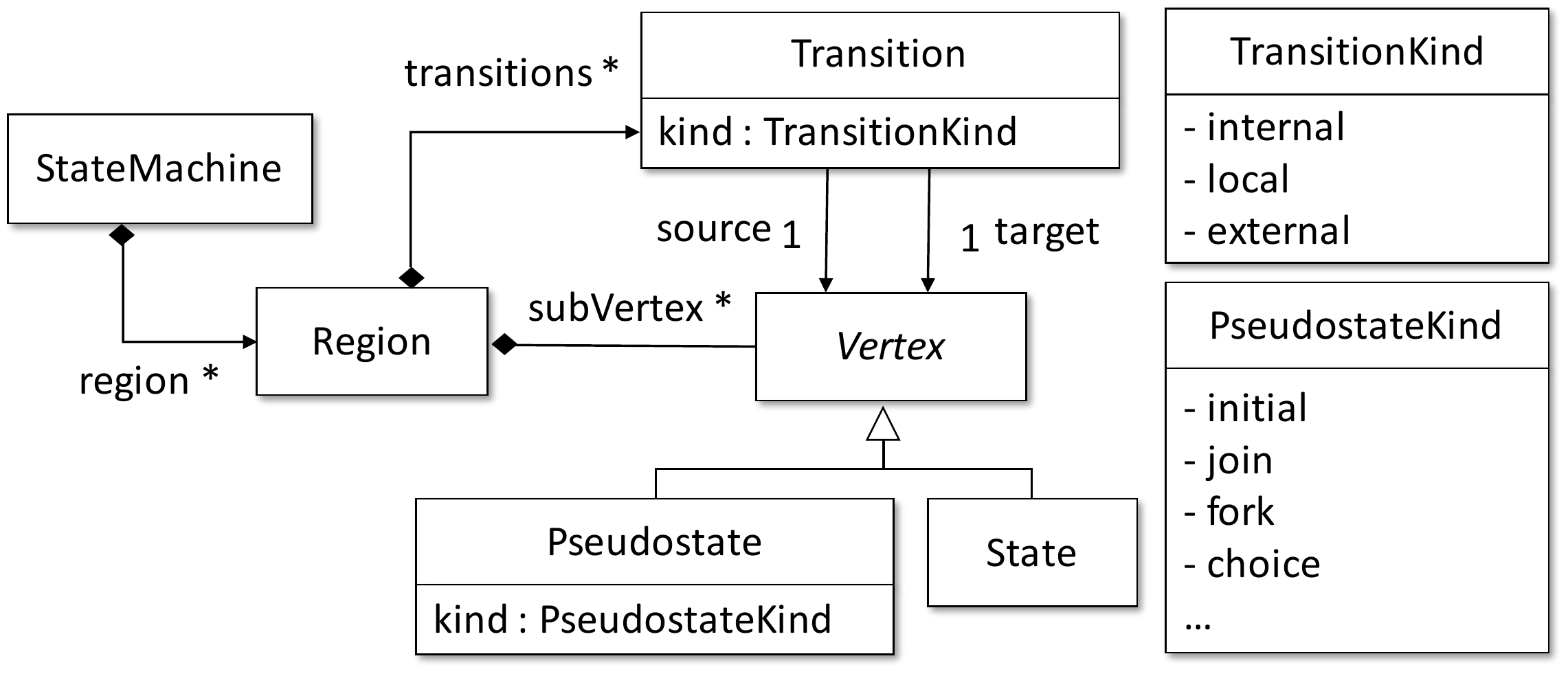}{0.5}{Excerpt of the UML state machine meta-model.}

\figpdf{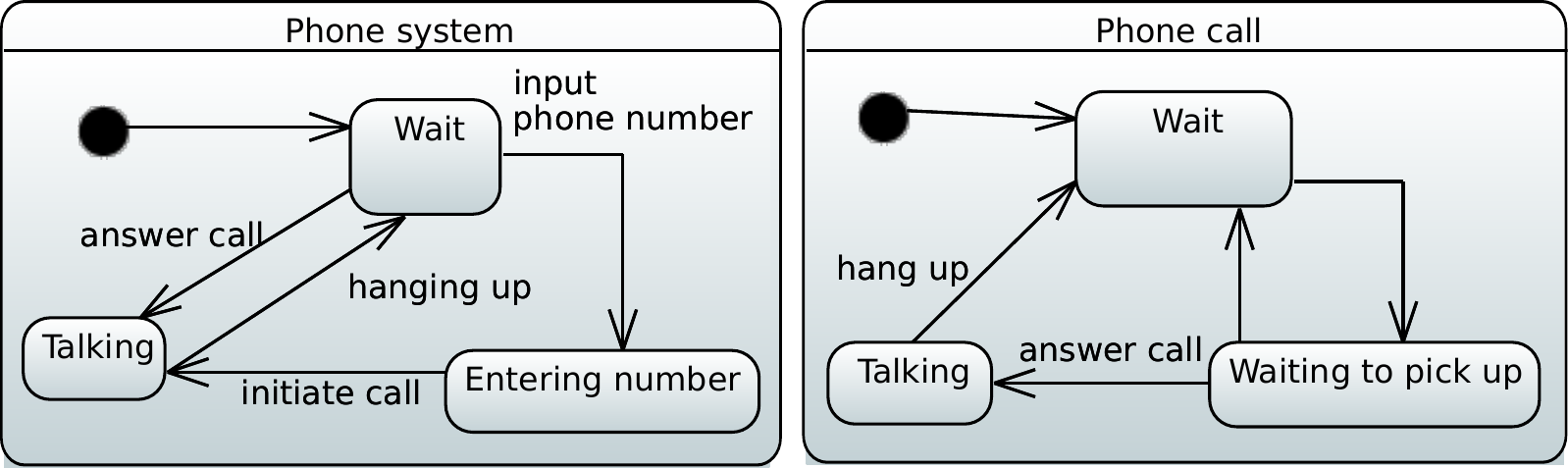}{0.5}{Repository model (left) and example query (right)}




\section{Retrieving models}\label{sec:model_retrieving}

Our approach is based on the conceptual model typically used to address text retrieval problems~\cite{robertson2009probabilistic},
but it is adapted to cover the specific case of model retrieval in which the structure of models
plays an important role. In this section, we present our approach formally, and the next
section describes the practical aspects.

\vspace{3pt}
\noindent\textbf{Model retrieval.} We define the model retrieval problem as follows: let $\mathcal{M} = \{m_1,\dots,m_t\}$ be a set of models which conform to the same meta-model and a query $q$ made by a user (which in our case is a model conforming to the same meta-model as the models in $\mathcal{M}$). We are interested in computing $R(q)\subset \mathcal{M}$ the set of relevant models to the user who defined the query $q$. However, it is not realistic to find exactly that set, so typically an approximation $R'(q)$ is considered. In particular, using a ranking approach, we have:
$$R'(q) = \{m \in \mathcal{M} |r(q,m) \geq \theta\},$$
where $r(\cdot,\cdot)$ is called the \textit{ranking function} and
$\theta$ is a threshold. In practice, the user is given a list of models sorted decreasingly by the scores of each model, and the threshold is implicitly defined by the user who browses the list in descending order until she or he considers appropriate. A good ranking function should rank all relevant documents on top of all non-relevant ones.

Our approach has two main ingredients, which are described in the rest of the section. First,
encoding the set $\mathcal{M}$ into a suitable type of document which represents the structure in a compact manner. 
We use the notion of graph paths to transform the repository into a set of {\em bags of paths}. Second,
choosing a scoring function to rank the models in the repository with respect to the query.


    



\subsection{Models as bags of paths}


A key element in our approach is the ability to perform fast, approximate structural matching.
As noted in~\cite{babur2017using}, taking the structure of models into account is essential
to precisely compare models in scenarios like search, clustering, clone detection, classification, etc.

\vspace{5pt}
{\noindent\bf Graph construction.}
Our approach is based on representing the structure of a model as a bag of paths.
To this end, we first define the structure of the graph from which paths will be extracted.

\begin{definition}

A \textit{directed multigraph} is a tuple $(V,E,f)$ where $V$ is a nonempty finite set whose elements are called vertices, $E$ is a finite set whose elements are called edges and $f:E \longrightarrow V \times V$ is a function which defines the source and target vertex of an edge.
The edges $e_1$ and $e_2$ are called \textit{multiple edges} (or multi-edges) if and only if $f(e_1)=f(e_2)$.

\end{definition}

\begin{definition}

A \textit{labeled directed multigraph} is a tuple of the form $G=(V,E,f,M_V,M_E)$ where

\begin{itemize}
    \item $(V,E,f)$ is a \textit{directed multigraph}.
    \item $M_V = \{\mu_V:V \longrightarrow L: 0<|L|<\infty\}$ is a set of \textit{vertex labeling functions}.
    \item  $M_E = \{\mu_E:E \longrightarrow L: 0<|L|<\infty\}$ is a set of \textit{edge labeling functions}.
\end{itemize}

\end{definition}

In particular we are interested in a labeled directed multigraph $G=(V,E,f,\{\mu^1_V,\mu^2_V\},\{\mu_E\})$ with three labeling functions: 
\begin{itemize}
    \item $\mu^1_V: V \longrightarrow \{\text{attribute},\text{class}\}$ which indicates if a vertex represents an attribute value or the class name of an object.
    \item $\mu^2_V: V \longrightarrow L_V = L_A \cup L_C$ maps vertices to a finite vocabulary which has two components:
    \begin{itemize}
        
        \item $L_A$ corresponds to the set of attribute values 
        in the model. 
        Given $v\in V$, if $\mu^1_V(v) = \text{attribute}$ then $\mu^2_V(v) \in L_A$.
        \item $L_C$ corresponds to the set of names of the different classes. Given $v\in V$, if $\mu^1_V(v) = \text{class}$ then $\mu^2_V(v) \in L_C$.
    \end{itemize}
    \item $\mu_E :E \longrightarrow L_E= L_R \cup L_{AR}$ maps edges to a finite vocabulary which has two components:
    \begin{itemize}
        \item $L_R$ corresponds to the set of names of the different references. Given $e\in E$ where $f(e) = (v,w)$, if $\mu^1_V(v) = \text{class}$ and $\mu^1_V(w) = \text{class}$ then $\mu_E(e)\in L_R$.
        \item $L_{AR}$ corresponds to the set of names of the different attributes. Given $e\in E$ where $f(e) = (v,w)$, if $\mu^1_V(v) = \text{class}$ and $\mu^1_V(w) = \text{attribute}$ or $\mu^1_V(w) = \text{class}$ and $\mu^1_V(v) = \text{attribute}$ then $\mu_E(e)\in L_{AR}$.  
    \end{itemize}
\end{itemize}

As can be noted, we explicitly consider two types of nodes: attribute
values and class names of individual objects (which will be referred to as ``object class'' nodes for brevity). The aim is to
be able to construct paths between the model elements which carry
the meaning of the model. For instance, the value {\em Waiting} of the \code{State.name} attribute would yield a node,
and each object of type \code{State} would yield another node
labeled as \code{State}.
The construction of the graph is relatively straightforward,
and we do not detail the algorithm for brevity.
Fig.~\ref{fig:multigraph} shows an excerpt of the graph
that is obtained for the query model shown in Fig.~\ref{fig:example_models}(right). 
There are explicit edges for both references and attributes.
Nodes associated
with objects classes are depicted as rounded rectangles, while nodes
associated with attributes are depicted as ovals. 

\figpdf{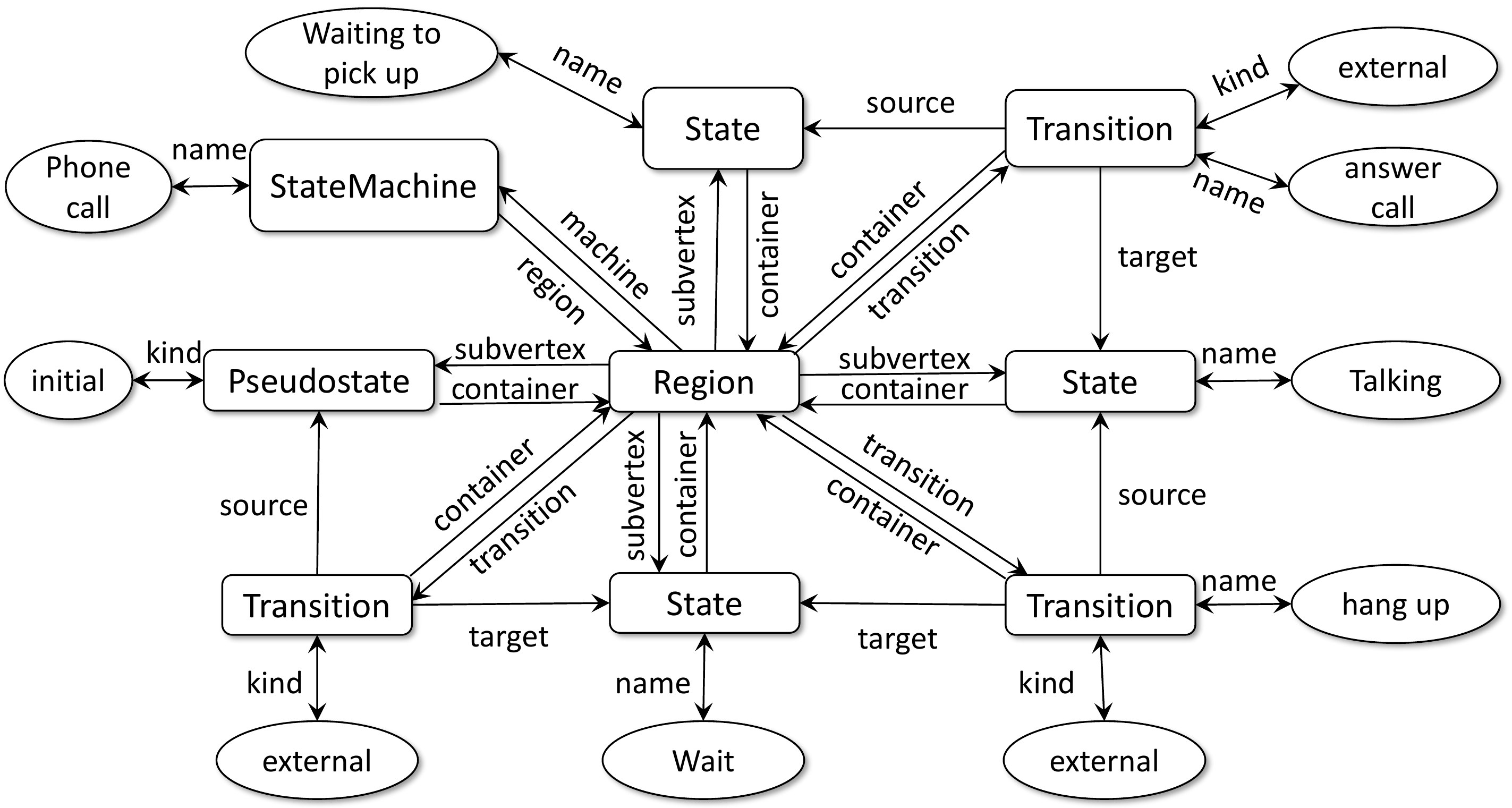}{0.5}{Excerpt of the multigraph for Fig.~\ref{fig:example_models}(right).}



\vspace{2pt}
{\noindent\bf Bag of paths.}
In our approach a bag of paths ({\em BoP}) refers to the paths between each
vertex of the multigraph to another one and the singleton paths with just one vertex. This is defined more formally 
in the following.


\begin{definition}

Let us have $G=(V,E,f,\{\mu^1_V,\mu^2_V\},\{\mu_E\})$. A \textit{bag of paths} for $G$ is a multiset $BoP$ whose elements are of the form:

\begin{itemize}
    \item $(\mu^2_V(v))$ where $v \in V$ or
    \item $(\mu^2_V(v_1),\mu_E(e_1),\dots,\mu^2_V(v_{n-1}),\mu_E(e_{n-1}),\mu^2_V(v_{n}))$ where
    \begin{itemize}
        \item $n \geq 2$
        \item $(v_1,\dots,v_n)$ a path of length $n-1$, i.e. a sequence of vertices where, for all $1\leq j\leq n-1$, $f(e_j) =(v_j,v_{j+1})$.
    \end{itemize}
\end{itemize}

\end{definition}




This definition of $BoP$ does not enforce which paths must be computed, but a criteria must be defined depending on the application domain.
For instance, the criteria could be to select all paths of the graph of length 3. In this case, the following paths\footnote{To facilitate reading the paths we use ovals to denote attribute nodes,  superscripted arrows to denote edges and rounded rectangles for class nodes.} would be part of
the $BoP$ of Fig.~\ref{fig:multigraph}: \cpath{\att{answer call}, \edge{name}, \cl{Transition}, \edge{target}, \cl{State}, \edge{name}, \att{Talking}}, \cpath{\cl{Transition}, \edge{container}, \cl{Region}, \edge{transition}, \cl{Transition}, \edge{source}, \cl{State}}, etc.
The concrete criteria that we use in the $BoP$ extraction for
our search engine is explained in Sect.~\ref{sec:path_extraction}. 

A $BoP$ has two main characteristics: it is a compact representation of the model structure and it can be used to reveal relevant models with respect to the query. For instance, the path \cpath{\att{answer call}, \edge{name}, \cl{Transition}, \edge{target}, \cl{State}, \edge{name}, \att{Talking}} extracted from the query graph encodes the fact that the user is interested models with a transition named \textit{answer call} whose target state is named \textit{Talking}. The same path exists in the repository model Fig.~\ref{fig:example_models}(left), hence this model become relevant to the query.



\subsection{Scoring and ranking models}\label{sec:scoring_ranking}



Given a query $q$, our goal is to determine the relevance of the models with respect to it.
To achieve this, the models that belong to $\mathcal{M}$ are transformed into a set of graphs. For each graph, the paths are extracted generating a set of bags of paths, $BoPs = \{BoP_1,\dots,BoP_t\}$, where $t = |\mathcal{M}|$ . On the other hand, the query $q$ is processed in a similar way generating the bag of paths $BoP_q$.






We are interested in comparing $ BoP_q$ with each $BoP_i$ in the repository. The comparison must take into account the number of
paths that appear in both $BoP_q$ and $BoP_i$ (matching paths) and the
relevance that each matching path has within each $BoP_i$. 
To this end, we have chosen the ranking function Okapi BM25~\cite{robertson2009probabilistic,Zhai2016TextDM} which is computed using the matching paths between both $BoP_q$ and $BoP_i$,
and taking into account the relevance of each match according to the size  $|BoP_i|$ in comparison to the sizes of the $BoP$s in the repository and the number of repetitions of each path in $BoP_i$.
Thus, $r\left(BoP_q,BoP_i\right)$ using our adapted
version of Okapi BM25 is:

\begin{equation}
\label{eq:okapi}
        \sum_{\omega \in BoP_q \cap BoP_i} \frac{c(\omega,BoP_q)(z+1)c(\omega,BoP_i)}{c(\omega,BoP_i)+z\left(1-b+b\frac{|BoP_i|}{avdl}\right)}\log\left(\frac{t+1}{df(\omega)}\right),
\end{equation}
where $1\leq i \leq t$, $c(\omega,BoP)$ is the number of times that a path $\omega$ appears in a bag $BoP$, $df(\omega)$ is the number of bags in $BoPs$ which have that path, $avdl$ is the average of the number of paths in all the $BoPs$ in the repository and $z \in [0,+\infty)$, $b \in [0,1]$ are hyperparameters. The parameter $b$ controls the penalization of models which have a large size and $z$ controls how quickly an increase in the path occurrence frequency results in path frequency saturation. If $b \rightarrow 1$, larger models have more penalization. In our current version we take $b=0.75$ as default value. If $z \rightarrow 0$ the saturation is quicker and if $z \rightarrow \infty$ the saturation is slower. We take $z=0.1$ (quick saturation) because a lot of repetitions of a path in a model does not imply that this model has much more relevance than other model that has this path repeated less times. 

Computing a ranking using this formula is straightforward, just 
by taking the query and all the models in the repository, computing $r(BoP_q,\cdot)$ for every model of the repository and sorting the models in decreasing order. However this approach is very inefficient since it has to visit all the models in the repository, performs the path extraction for each one and compute $r$. Next section describes how to engineer the search engine to efficiently 
compute the ranking.

\section{Engineering a search engine}\label{sec:components}



In this section, we describe the practical aspects of our search engine.
The architecture is depicted in Fig.~\ref{fig:offlineonline}. There are
two main processes: offline and online. The offline process is in charge
of gathering models from one or more model repositories using
dedicated {\em crawlers}. The indexer takes these models, process them, and 
constructs an inverted index over Apache HBase using a specific encoding
to make the search fast. For each model, a pipeline
made of several components is applied: a {\em model to graph} transformation, a {\em path extraction} process to 
obtain a bag of paths of a given length from the graph (see Sect.~\ref{sec:path_extraction}), a {\em normalization}
process in which some standard Natural Language Processing (NLP) techniques are applied to have a more semantic handling of string values (see Sect.~\ref{sec:normalization}), and the {\em stop path} removal process which filters out some irrelevant paths from the bag of paths. 

On the other hand, given a query we apply
the same pipeline. In this case, the resulting bag of paths is used
by the scorer, which implements the scoring function in Equation~\ref{eq:okapi} by accessing the index, and returns a ranked list of models according to their similarity.

In the following we describe the design and some implementation details
of these components.

\figpdf{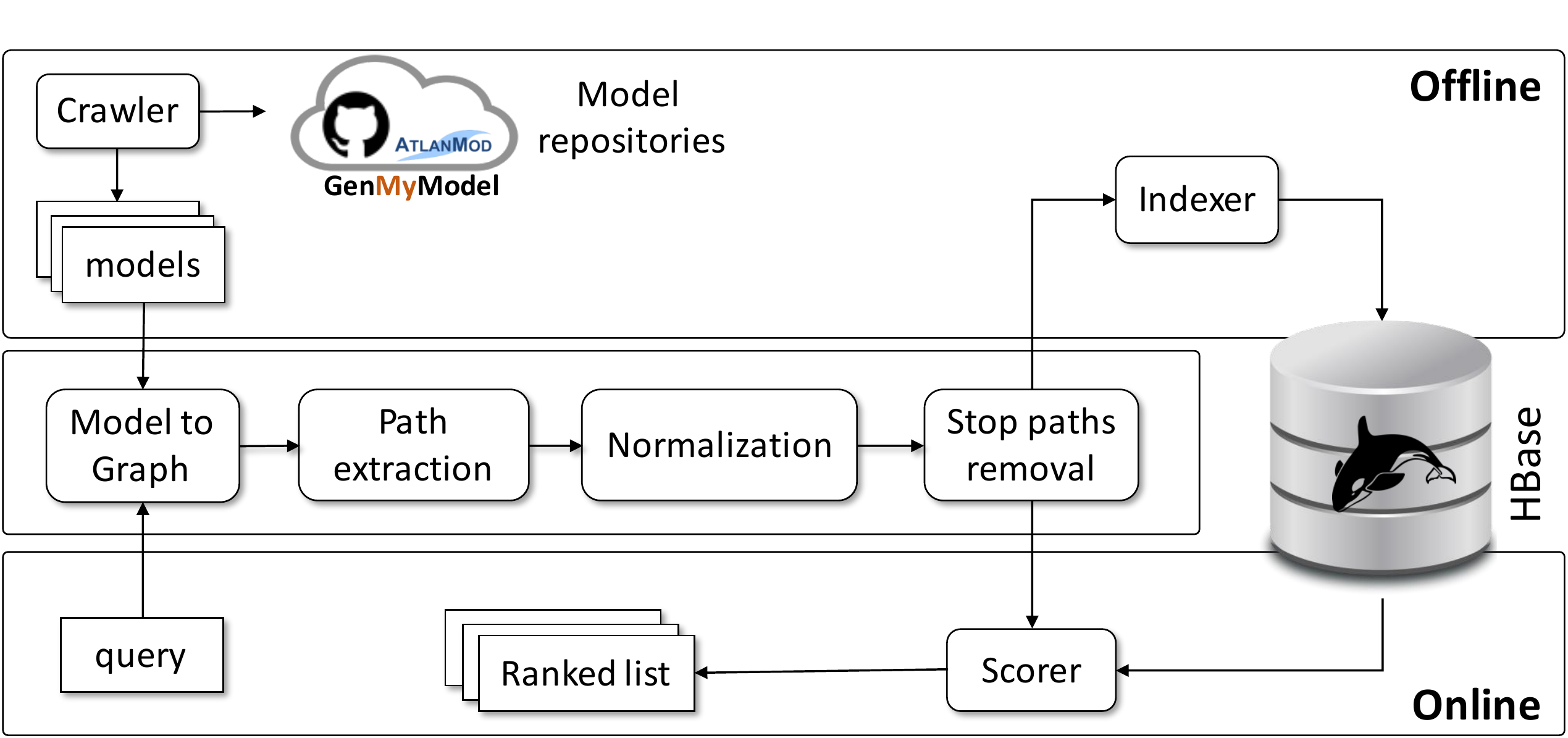}{0.5}{Architecture and main components of MAR.}

\subsection{Path extraction}\label{sec:path_extraction}


Path extraction is an important element of our search engine, since it transforms the structural information of the
model into a form suitable for indexing. In the previous section we define the multiset $BoP$ by describing the form of the elements that belong to it in a generic way. In practice, we need to
define a criteria to extract a relevant $BoP$ from a model. 
To establish this criteria we follow this reasoning: 
increasing the length of the path is equivalent to increasing the requirements that a model of the repository has to satisfy in order to be relevant to the user query. For instance, for the graph in Fig.~\ref{fig:multigraph} we can extract these two paths:
\cpath{\att{answer call}, \edge{name}, \cl{Transition}} and \cpath{\att{answer call}, \edge{name}, \cl{Transition}, \edge{target}, \cl{State}, \edge{name}, \att{Talking}}. The first path (length one) is less restrictive in the sense that all models which have a transition called \textit{answer call} will be relevant. The second path (length three) is more restrictive since it states that relevant models need to have a transition called \textit{answer call} that has a target state of name \textit{Talking}. 
Therefore, we construct the $BoP$ by considering all simple paths (no cycles) of length {\em less or equal} than a threshold (typically 4). On the other hand, even with a fixed length, considering all paths between all vertices will make a huge $BoP$. 
Consequently, we consider the paths with the following forms:
\begin{itemize}
    \item All singleton paths for objects with no attributes: $(\mu^2_V(v))$ where $v \in V$, $\mu^1_V(v) = \text{class}$ and $\forall w$ accessible from $v$, $\mu^1(w)=\text{class}$. Path \cpath{\cl{Region}} is the only one in the example.
    
    \item All paths of $length=1$ between attribute values and its associated object class: $(\mu^2_V(v),\mu_E(e),\mu^2_V(u))$ where $\mu^1_V(v) = \text{attribute}$. These encode object data. For instance, \cpath{\att{initial}, \edge{kind}, \cl{PseudoState}} and \cpath{\att{Phone call}, \edge{name}, \cl{StateMachine}}.
    
    \item All paths without cycles between attributes; and all paths without cycles between attributes and objects without attributes: $(\mu^2_V(v_1),\mu_E(e_1),\dots,\mu^2_V(v_{n-1}),\mu_E(e_{n-1}),\mu^2_V(v_{n}))$ \linebreak where $n - 1 \leq length~threshold$ and one and only one of the following statements is true:
    \begin{itemize}
        \item $\mu^1_V(v_1) = \text{attribute}$ and $\mu^1_V(v_n) = \text{attribute}$.
        \item $\mu^1_V(v_1) = \text{attribute}$ and $\mu^1_V(v_n) = \text{class}$ where $\forall w$ accessible from $v_n$, $\mu^1(w)=\text{class}$.
        \item $\mu^1_V(v_n) = \text{attribute}$ and $\mu^1_V(v_1) = \text{class}$ where $\forall w$ accessible from $v_1$, $\mu^1(w)=\text{class}$.
        \item $\mu^1_V(v_1) = \text{class}$ where $\forall w$ accessible from $v_1$, $\mu^1(w)=\text{class}$. $\mu^1_V(v_n) = \text{class}$ where $\forall w$ accessible from $v_n$, $\mu^1(w)=\text{class}$.
    \end{itemize}
    For instance, with $length\leq4$ we have paths like:
    \begin{itemize}
        \item \cpath{\att{answer call},\edge{name},\cl{Transition},\edge{kind},\att{external}} 
        \item \cpath{\att{Phone call}, \edge{name}, \cl{StateMachine}, \edge{region}, \cl{Region}}
        \item \cpath{\att{external},\edge{kind},\cl{Transition}, \edge{source}, \cl{State},\edge{name},\att{Wait to pick up}}
        \item \cpath{\att{Talking},\edge{name},\cl{State}, \edge{container}, \cl{Region},\edge{subvertex},\att{State}, \edge{name}, \att{Phone call}}
    \end{itemize}
\end{itemize}

The rationale is to represent the data of an object (its internal state) with 
paths of $length=1$, or with $length=0$ if an object does not have attributes
and we use the class name as a surrogate. We use paths of $length > 1$ to
encode the model structure in terms of the relationships between the data of the objects (their attribute values).
In practice, \mar is configurable to reduce the $BoP$ size
even more by filtering out
classes, attributes or references which might not
be of interest for specific meta-models.
    



    


\subsection{Normalization}\label{sec:normalization}

Once the $BoP$ of a model has been computed, the next step is to normalize to increase the chances of having matching paths.
\subsubsection{Word normalization}
Oftentimes attribute values are strings, which typically carry some meaning according to the model domain. However, small variations in the same word prevent matching similar words.
Given an attribute of type string, we apply the following
techniques typically used in Natural Language Processing (NLP):


\begin{itemize}
    \item Tokenization: Given a character sequence, tokenization is the task of dividing it into pieces, called tokens. This process is customizable. 
    In the running example we use a \textit{white space tokenizer}
    and make characters lower case. The tokenization of  \textit{Waiting to pick up} results in \textit{[waiting] [to] [pick] [up]}. 
    
    \item Stop words removal: The term stop word refers to a word that appears very frequently when using
    a language. Thus, a typical process in NLP systems is to remove them. In our case, we remove tokens
    containing English stop words. In the previous example we would apply stop word removal as follows: \textit{[waiting] [to] [pick] [up]} $\longrightarrow$ \textit{[waiting] [pick]}.
    
    \item Stemming: This is a NLP technique whose goal is to reduce inflectional forms and sometimes derivationally related forms of a word to a common base form. In particular we use the Porter Stemming Algorithm \cite{Porter1980AnAF}.
    In the example, the application of stemming would produces: 
    \textit{[stem(waiting)] [stem(pick)]} $\longrightarrow$ \textit{[wait] [pick]}.
\end{itemize}

At the end of this process we might end up with a single string value (e.g., ``Waiting to pick up``) split into zero or more tokens. If there are no tokens, the node is removed. If there are more than one token, the node is duplicated according to the split and all the paths including such node are duplicated accordingly. For instance, the path \cpath{\att{Waiting to pick up}, \edge{name}, \cl{State}} is transformed in these two paths: \cpath{\att{wait}, \edge{name}, \cl{State}} and \cpath{\att{pick}, \edge{name}, \cl{State}}.




\subsubsection{Stop paths removal}
There exists some paths which do not provide any information about the model, in the sense of how similar or different is from other models. This happens for paths which appear in the majority of the models, which we call \textit{stop paths}. 
For example, a path like \cpath{\att{initial},\edge{kind},\cl{PseudoState}} is a stop path because most state machines have an initial state. 
Currently, we heuristically consider a path to be a stop path if it appears in the $70\%$ of the models in the repository. This is calculated only once as a post-processing in the indexing.

\subsection{Indexing models}\label{sec:indexer}

In a search engine, the indexer is in charge of organising the documents in a way that enables
a fast response to the queries. In our case, the documents indexed by the
search engine are models obtained by crawling existing model repositories.
We have created a number of scripts to semi-automate
this process for GitHub, GenMyModel and the AtlanMod Zoo. The results are
collections of XMI files which are consumed by the indexer.

The main data structure used by the indexer is the {\em inverted index}~\cite{arasu2001searching}.
It is a large array with one entry per word in the global vocabulary and each entry points to a list of documents that contains such word. In \mar, we index models, and therefore there is an entry per different
path and, for each one, a pointer to the list of models which have this path. However, compared to text retrieval systems the size of the index is larger since there are much more paths than words and queries are bigger (in a text retrieval system the queries are keywords), that is, there are more accesses to the inverted index. For instance, in a repository of 17.000 models, we counted more than twenty million different paths and a small query could have hundreds of different paths. 

We use Apache HBase~\cite{george2011hbase} to implement the inverted index. HBase is a sparse, distributed, persistent multidimensional sorted map database, inspired by Google's BigTable~\cite{chang2008bigtable}. 
HBase is prepared to provide random access over large amounts of information, and its
horizontally scalable, meaning that it scales just by adding more machines into the pool of resources.
In HBase each
table has rows identified by row keys, each row has some columns (qualifiers) and each column belongs to a column family and has an associated value. A distinctive feature of HBase is that two rows can have different columns, and a row may have thousands of columns. 
HBase provides two read operations: \textit{scan} and \textit{get}. \textit{Scan} is used to iterate over all rows that satisfy a given filter. \textit{Get} is used for random access by key to a particular row.

Building an inverted index over HBase can be straightforward if each path is a row key and the models that contain this path are columns
whose qualifier (i.e., the column name) is the model identifier and whose value is a pair containing the number of times the models have the path and the total number of paths of the model (this information is pre-computed
to calculate the score efficiently). 
However, this approach is not optimal because to implement the scoring
function we need to perform as
many {\em get} calls as paths in the query. This causes a performance bottleneck because 
each get is completely independent; that is, for each path in the query (e.g., a thousand paths) an independent look up is done just to recover the needed information about one and only one path. Another option is to use a \textit{scan} with a prefix filter. However, this can return a lot of paths that we have to filter in the client side.

An alternative is to organize the table schema in a way that 
accommodates to the most common access pattern, so that 
the data that is read together is stored together. In HBase the data is stored following a lexicographic order of the concatenation of the row key and the column qualifier in a column family. 
Therefore, we propose a database schema similar to the one in 
Fig.~\ref{fig:CFSchema1}.
Each type of model has an inverted index (a table). 
Each path is split into two parts: the prefix and the rest, and
the prefix is used as row key and the rest is a column qualifier. The split is different depending on whether it starts with a attribute
node or an object class node.

\vspace{2pt}
{\noindent \bf Path starts with attribute value}. The prefix is
the first sub-path of length 1. For instance, the path \cpath{\att{hang}, \edge{name}, \cl{Transition}, \edge{source}, \cl{State}, \edge{name}, \att{talk}} is split into a row key: \textit{(hang, name, Transition} and the column qualifier: \textit{, source, State, name, talk)}. The value associated to the column is a serialized map which associates the identifier of the models which have this path with the number of times the path appears in the models and the total number of paths of the models (these two numbers are necessary in the calculus of the score). In the figure, \code{sm1: [1, 1032]} means that the path appears in model \code{sm1} only once and \code{1032} is the number of paths that \code{sm1} has.
For paths of $length=1$ we use the symbol $)$ as a marker for the column
qualifier, as in \cpath{\att{hang}, \edge{name}, \cl{Transition}}.
        
\vspace{2pt}
{\noindent \bf Path starts with object class}. For instance, an example could be \cpath{\cl{Region}, \edge{subvertex}, \cl{State}, \edge{name}, \att{talk}}. This path is split into \textit{(Region} and \textit{, subvertex, State, name, talk)}. \textit{(Region} is going to be the row key and  \textit{, subvertex, State, name, talk)} will be the column qualifier. 
For paths of null length, for instance \cpath{\cl{Region}} we use
$)$ as column name as before.


\figpng{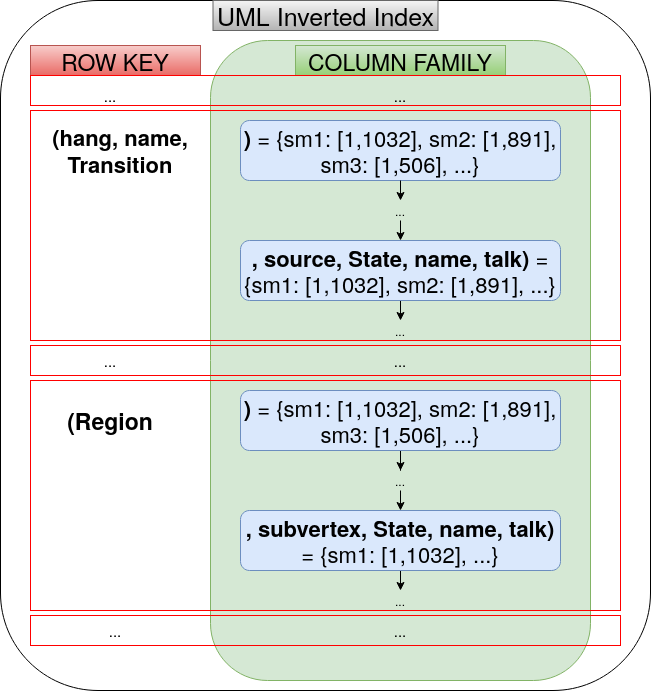}{0.4}{HBase table schema}

\subsection{Scorer}


%
%
%
%


Using this schema, the underlying idea to implement an optimized version of the
scoring function is to gather 
information about more than one path in each database $get$ petition. Given a query, for each distinct prefix there is only one petition to HBase because it is possible add the columns that you want to retrieve in a single \textit{get}. The main idea is to explode the lexicographic order in each petition returning the needed information about some paths that are stored together due to the fact that they have the same row key (i.e. the same prefix).
The following algorithm shows how to implement the scoring function
for HBase.


\begin{algorithm}
 \KwData{$BoP_q$ = Bag of paths of the query, $avdl$ = average of path lengths, $t$ = number of indexed models.}
 \KwResult{scores = map of <model, score>}
 
  \ForAll{distinct prefix $p$ of paths $BoP_q$}{
  
  
  
  //For every prefix extract the corresponding rests
  
  $cols \leftarrow \text{ distinct elements of } \left\{x | \omega = px \text{ and } \omega \in BoP_q\right\}$
  
  
  
  $results \longleftarrow$ get(row=$p$,columns=$cols$);
  
  \ForAll{row key, column, models $\in$ results}{
 
  $\omega \longleftarrow$ concatenation(row key, column);
        
  $c(\omega,BoP_q)\longleftarrow$ $BoP_q[\omega]$;
        
  $df(\omega)\longleftarrow$size(models);
        
  \ForAll{$m$,$c(\omega,BoP_m),|BoP_m|\in$ models}{
 
            update scores$[m]$ using $c(\omega,BoP_m)$, $c(\omega,BoP_q)$, $avdl$, $t$, $|BoP_m|$, $df(\omega)$ and the equation (\ref{eq:okapi});
  }
  
  }
  
  }
\end{algorithm}

In this algorithm paths are split into prefix and rest as explained in Sect.~\ref{sec:indexer}, and elements $\omega$, $c$ and $df$ were explained in Sect.~\ref{sec:scoring_ranking}.
If a user introduces the query Fig.~\ref{fig:example_models}(right) in our search engine, it would return a ranked list in which the first result is the one shown in Fig.~\ref{fig:example_models}(left) with a score of $1520.15$. Some of the paths that contribute to the score (i.e. that match) are \cpath{\att{wait}, \edge{name}, \cl{State}}, \cpath{\att{talk}, \edge{name}, \cl{State}}, \cpath{\att{answer}, \edge{name}, \cl{Transition}, \edge{target}, \cl{State}, \edge{name}, \att{talk}}, etc.




\section{Applications}\label{sec:applications}

The main application of our search engine is obviously searching for models, but it can also be 
applied to other scenarios. 
For instance,  it can be used as the basis to build a model recommender
which searches in the background for relevant models and then extracts
new features to suggest (e.g., a property for a class). It can also be used
as a means to find reusable MDE components (e.g., transformations, editors, etc.) by indexing the meta-model footprint of each component. 
In this section we discuss two applications for which we have
already used and experimented with our search engine:
searching for models and meta-model classification, but
we want to implement new applications in the future.

\vspace{-5pt}
\subsection{Model searching}

The availability of model repositories has been regarded as an
important element for the success of MDE, since they may foster
the reuse and sharing of software models, thus increasing the visibility
of the MDE paradigm. There are several model repositories (e.g., ReMoDD, AtlanMod Zoo, GenMyModel), but models are also stored in regular code repositories 
(e.g., as part of GitHub projects). This makes it difficult for users
to retrieve relevant models, since it implies looking up in several locations. Moreover search tools in these repositories are not model-oriented or there might be no search tool at all~\cite{di2015collaborative}.
\mar is able to aggregate models coming from several
sources and provides an unified access to them. 
\mar is available at \url{http://mar-search.org},
in which updated statistics about the number and types of models
can also be checked. We also
provide access to \mar through a REST API and an Eclipse-plugin.


\vspace{3pt}
{\bf \noindent Access through a REST API.}
We have implemented a simple REST API to allow the integration of
the search engine in tools of diverse nature, like a web-based
interface, extensions of modelling environments, or simply to perform experiments.

The REST API has two main methods: \code{search} is a POST method which
receives a file in XMI format, the type of model and an integer with 
the maximum number of results. It connects with
the search engine and returns as a response a JSON document with the list
of relevant document as pairs of (id, score). The GET method \code{get\_model} retrieves the data associated to a given model. 

\vspace{3pt}
{\bf \noindent Eclipse plug-in.} As a prototype we have implemented 
an Eclipse plug-in to add search functionality to existing Eclipse 
model editors. Fig.~\ref{fig:screenshot} shows a screenshot. An Eclipse
view connects to the active editor. The {\em search}
button takes the contents of the current editor, serializes it and 
submits the query. The results can be inspected using a dedicated view
or downloaded for further manipulation.

\figpdf{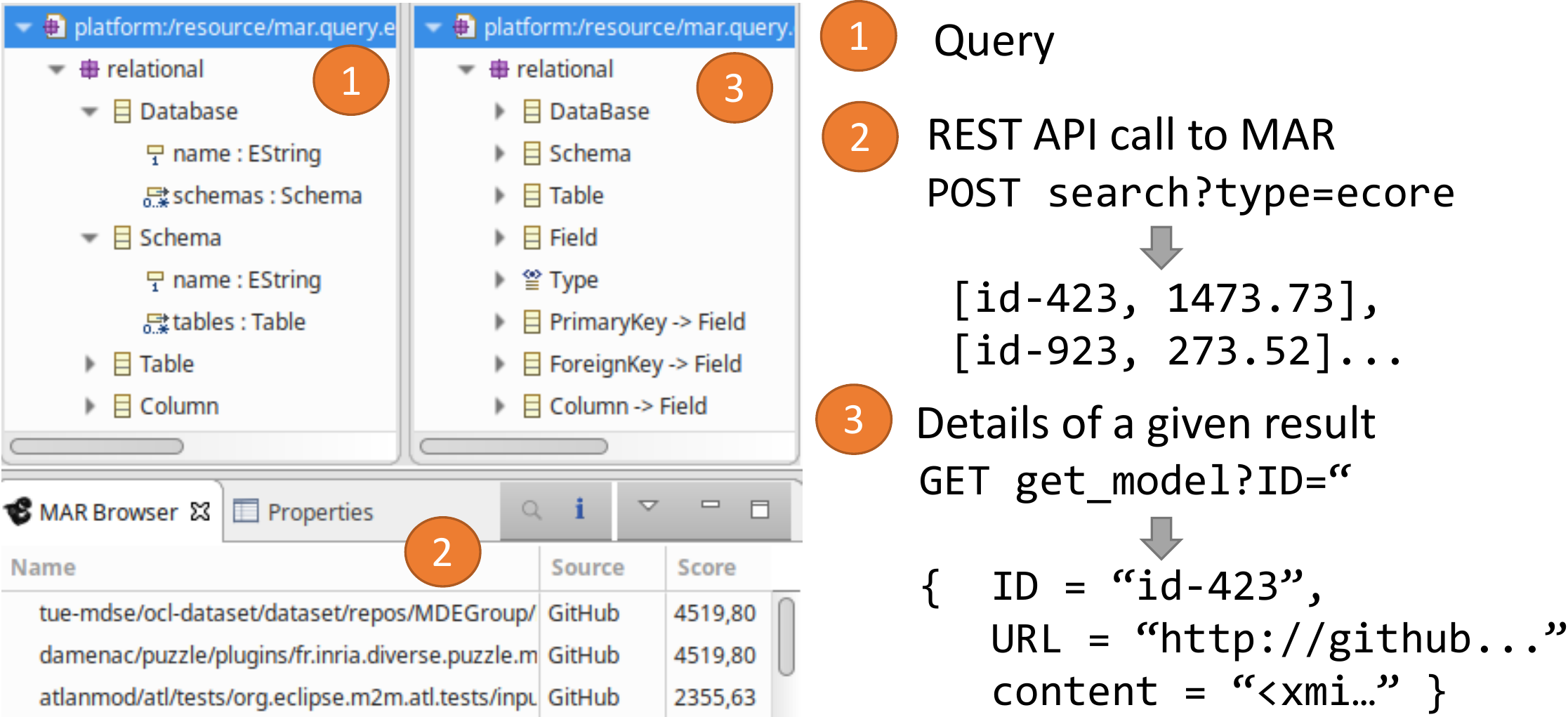}{0.5}{Eclipse-plugin and REST API}




\subsection{Meta-model classification}

The automated classification of model repositories is regarded as
an important activity to support reuse in large model repositories~\cite{nguyen2019automated}.
Automated methods could annotate models with fine-grained metadata to help developers make decisions about what to reuse.

Recent works have used supervised machine learning methods like neural networks~\cite{nguyen2019automated} and 
unsupervised methods like clustering~\cite{babur2016hierarchical,basciani2016automated} to
classify meta-models into application domains automatically
(e.g., state machines, class diagrams, etc.).
An alternative and simpler method is to use a model search engine like ours to apply the
$k-$Nearest Neighbors ($k-$NN) supervised learning algorithm to this task~\cite{Zhai2016TextDM}. 
Let us suppose that we have a repository labeled with the 
domain of each meta-model. If we want to classify a new meta-model,
we use it as a query for our search engine, we select the top $k$ meta-models of the ranked list of results, and finally we assign to it the label of the majority. 


 
To take advantage of the score provided by the search engine we use weighted $k-$NN, which is a modification of $k-$NN. Weighted $k-$NN is based on assigning weights to the votes of the test data's neighbors. In our case, the weights will be the score of each result.


To evaluate the adequacy of this approach we replicate the 
experiment carried out in~\cite{nguyen2019automated} and compare
the results. We use the same dataset of 555 metamodels \cite{onder_babur_2019_2585456} retrieved from GitHub. In this dataset,
each meta-model is labeled with its domain (9 domains were identified).
We follow this procedure to estimate the performance of weighted $k-$NN using \mar.

\vspace{1pt}
{\bf\noindent Selecting $\pmb k$.}
Split the 555 metamodels into two sets: training set ($70\%$ of them) and test set ($30\%$ of them). We use the training set to select the $k$ hyperparameter (i.e., number of neighbours) and estimate the performance. We do a $10-$fold cross-validation and, as evaluation metric, we use the accuracy (i.e., the percentage of correctly classified models). After the $10-$fold cross-validation, for each considered $k=2\dots 10$, we choose the $k$ that maximizes the accuracy's mean of validation sets.
The best results are achieved with $k = 2$ with an accuracy of $93.23\%$.



\vspace{1pt}
{\bf\noindent Test.} Once the $k$ hyperparameter has been chosen, 
we can directly evaluate the test set. In particular, with 
$k=2$ the accuracy of our method in the test set is $93.37\%$.



{\bf\noindent Assessment.}
The approach in~\cite{nguyen2019automated} uses a neural network
and estimate an accuracy of $95.45\%$, which is slightly better
than ours ($93.23\%$). Hence, it can be claimed that \mar can
also be used for meta-model classification with results comparable
to state-of-the-art techniques.
It is also worth mentioning that neural networks are very powerful but they are black box, that is, you do not know the reason of its predictions. In this way, our approach is more transparent because the classification decision is determined easily by just looking at the models of the ranked list of results.






\section{Evaluation}\label{sec:evaluation}

    
This section reports the results of the evaluation of our search engine.
First, we evaluate its search precision, that is, the ability of \mar to rank
relevant models on top. Second, we show the performance results.
The datasets and a replication package are available at \url{http://mar-search.org/experiments/models20}.


\subsection{Search precision}
The evaluation of the precision of a search engine requires
some metric about the adequacy and relevance of the search results. A user-oriented study about
the relevance of the search results would be an appropriate manner to evaluate the usefulness of our system.
However, setting up this kind of study is out of the scope of the paper and it is left as future work. Instead,
to have an initial evaluation of \mar we have devised an automated approach.



The underlying idea is to simulate a user who has in mind a concrete
model and the query is automatically derived using mutation operators
that generate a shrinked version of the model which includes some structural and naming changes. Thus, for each generated query mutant there is only one relevant model, which known beforehand. This type of evaluation is called \textit{known item search}~\cite{Zhai2016TextDM}.

As target for the search, we have used all the Ecore meta-models indexed by \mar (17975). 
We perform the evaluation twice, with two different sets of mutants. The first set comes from 281 meta-models obtained from 
the AtlanMod zoo (referred to as $\text{Mutants}_\text{AtlanMod}$).
The second set comes from 2000 meta-models randomly selected from all the
indexed meta-models (referred to as $\text{Mutants}_\text{All}$).
To ensure a certain degree of quality in the query mutants, we require that the
meta-models have at least 20 classes and 40 elements (adding up \code{EClass} and \code{EStructuralFeature} elements). The aim is to filter out small meta-models which may result in queries very close to the original.

For each
meta-model we apply in turn the mutation operators summarized in Table~\ref{tab:mutants}. First, we identify a potential root 
class for the meta-model and we keep all classes within a given ``radius'' (i.e., the number of reference or supertype/subtype relationships which needs to be traversed to reach a class from the root), and the rest are removed. All \code{EPackage} elements are renamed to avoid any naming bias. From this, we apply mutants to
remove some elements (mutations 2--6). Next, mutation \#7 is intended to make the query more general
by removing elements whose name is
very specific of this meta-model and it is almost never used in other meta-models 
(in text retrieval terms the element name has a low document frequency). Finally, to implement mutation \#8 we have 
applied clustering (k-means), based on the names of the elements of the meta-models, in order to group meta-models that belong to the same domain. In this way, this mutant attempts to apply meaningful renamings by picking up names from other meta-models
within the same cluster. We apply this process with different
radius configurations (5, 6 and 7) and we discard mutants
with less than 3 classes or with less references than $|classes| / 2$.
Using this strategy we generate 128 queries for $\text{Mutants}_\text{AtlanMod}$ and  1595 queries for $\text{Mutants}_\text{All}$.



Given that we do not have access to any other model search engine,
we have implemented a text-based
model search engine on top of Whoosh~\cite{Whoosh} (a pure Python search engine library) as way to have a baseline to compare against. The full repository is translated to text documents which contain the names of each meta-model (i.e., property \code{ENamedElement.name}). These documents are indexed and we associate the original meta-model to the document. Similarly, we generate the text counterparts of the mutant queries. 

\begin{table}[t]
\footnotesize
\begin{tabular}{rlp{5cm}}
\hline
& \textbf{Mutant}           & \textbf{Description}                                                    \\ \hline
1 & Extract connected subset        & Select a root element and picks classes reachable via references or subtype/supertype relationships, up to a given length.                                                                       \\ \hline
2 & Remove inheritance         & Remove a random inheritance link (up to 20\%).                                                              \\ \hline
3 & Remove leaf classes        & Remove classes, but prioritize those which are
farther from the root element (up to 30\%)            \\ \hline
4 & Remove references          & Remove a random reference (up to 30\%). 
                                            \\ \hline
5 & Remove enumeration         & Remove random enumerations or literals (50\%).                              \\ \hline
6 & Remove attributes           & Remove a random attribute (up to 30\%).                     \\ \hline
7 & Remove low-df classes      & Remove elements whose name is "rare".                     \\ \hline
8 & Rename from cluster        & Replace a name by another name corresponding to element belonging to the same meta-model cluster (up to 30\% of names). \\ \hline
\end{tabular}
\caption{Mutations operators used to simulate queries.}
\label{tab:mutants}
\end{table}

\begin{table}[t]
\footnotesize
\begin{tabular}{l|l|l|}
\cline{2-3}
                                         & $\text{Mutants}_\text{All}$         & $\text{Mutants}_\text{AtlanMod}$    \\ \hline
\multicolumn{1}{|l|}{MAR -- MRR}          & 0.752                            & 0.968                            \\ \hline
\multicolumn{1}{|l|}{Whoosh -- MRR}       & 0.668                            & 0.894                            \\ \hline
\multicolumn{1}{|l|}{Differences in MRR} & +0,084                           & +0,074                           \\ \hline
\multicolumn{1}{|l|}{$p-value$}          & <.001 & <.001 \\ \hline
\end{tabular}
\caption{Results of search precision evaluation.}
\label{tab:mrr}
\end{table}

The procedure to evaluate the precision of each search engine and query set is as follows. Each query has associated the meta-model from which it was derived. For each query we perform a search and we retrieve the ranked list of results. We look up the original meta-model in the list and take its position, $r$. Then, we
compute the reciprocal rank which is $\frac{1}{r}$.
To summarize all reciprocal ranks we take the average of them (Mean Reciprocal Rank, MRR).
To compare the precision of \mar and Whoosh we use the parametric $t-$test for the two set of queries. In both cases, we have $p-value<.001$ indicating that the difference in the mean reciprocal rank is significant.
Table~\ref{tab:mrr} shows the global results of the evaluation and
Fig.~\ref{fig:marvswhoosh} shows the proportion of queries which are ranked in the first, second up to the fifth position or more.
\mar ranks more queries in the first position in both query
sets. 
It is also interesting to observe that the proportion of queries
ranked in positions equal or greater than 5 is smaller in \mar.  
In addition, the results for $\text{Repo}_\text{AtlanMod}$ are better than
$\text{Repo}_\text{All}$ in both \mar and Whoosh. The reason is that the AtlanMod meta-models are generally more complete (e.g., more classes, features, etc.), than the random sample taken from \mar's index. Therefore the query mutants tend to be of better quality.


From these results we conclude that \mar has a good precision,
and the precision seems to improve when we search for larger
models using well defined queries. 
The fact that \mar is better in both query sets reinforces
our hypothesis that \mar outperforms plain text search in general.



\figpng{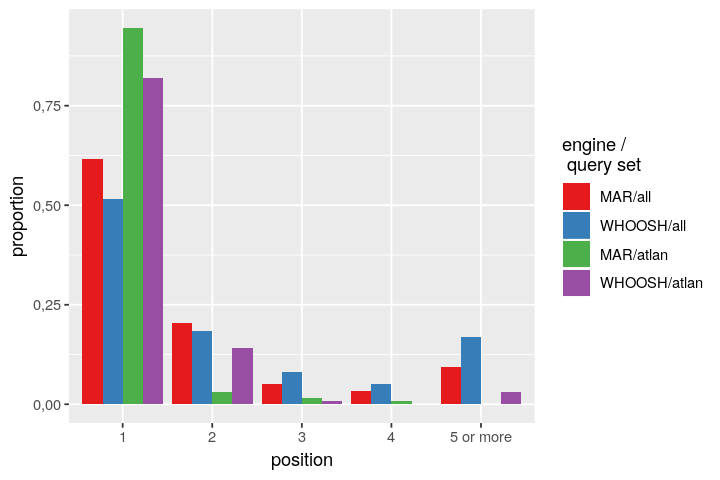}{0.5}{Precision evaluation results. Proportion of queries (y-axis), the position of the relevant document (x-axis) in the ranked list and engine/query set as colour of the bars.}

The main threat to validity of this experiment is that the obtained mutants might not represent
the queries that an actual user would create. To minimise it we have tried several 
mutation operators and configurations and reviewed the mutants manually until we found
reasonable results. Nevertheless, we expect to understand better the user behaviour
as \mar is used by the community.


\begin{table*}[t]
\centering
\footnotesize
\setlength{\tabcolsep}{4pt}
\begin{tabular}{cc|cccc|cccc|cccc|cccc|c|}
\cline{3-19}
                                        &           & \multicolumn{4}{c|}{EPackage} & \multicolumn{4}{c|}{EClass} & \multicolumn{4}{c|}{EAttribute} & \multicolumn{4}{c|}{EReference} & Total   \\ \cline{3-18}
                                        &           & min  & median  & mean  & max  & min  & median & mean  & max & min   & median   & mean  & max  & min   & median  & mean   & max  & queries \\ \hline
\multicolumn{2}{|c|}{$\text{Mutants}_\text{AtlanMod}$ (128 queries)} & 1    & 1       & 2.8   & 46   & 8    & 20     & 29.9  & 125 & 0     & 13       & 18.9  & 90   & 4     & 17      & 24.93  & 97   & 128     \\ \hline
\multicolumn{2}{|c|}{$\text{Mutants}_\text{All}$ (1595 queries)}     & 1    & 1       & 2.1   & 46   & 3    & 13     & 17.8  & 193 & 0     & 8        & 14.9  & 168  & 2     & 10      & 14.5   & 116  & 1595    \\ \hline
\multicolumn{1}{|c|}{$\text{Mutants}_\text{All}$}        & small     & 1    & 1       & 1.2   & 8    & 3    & 5      & 5.3   & 11  & 0     & 2        & 2.4   & 11   & 2     & 4       & 3.8    & 11   & 337     \\ \cline{2-19} 
\multicolumn{1}{|c|}{(1595 queries)}    & medium    & 1    & 1       & 1.6   & 18   & 3    & 13     & 13.4  & 35  & 0     & 8        & 9.8   & 45   & 2     & 10      & 10.8   & 29   & 848     \\ \cline{2-19} 
\multicolumn{1}{|c|}{}                  & large     & 1    & 1       & 3.6   & 46   & 3    & 27     & 37.25 & 193 & 1     & 29       & 35.7  & 168  & 2     & 24      & 30.9   & 116  & 410     \\ \hline
\end{tabular}
\caption{Statistics of the datasets used in the evaluation.}
\label{tab:stat}
\end{table*}

\subsection{Performance evaluation}
To evaluate the performance of \mar we have used all mutant queries extracted from $\text{Mutants}_\text{All}$. We want to
evaluate the effect of the index size in the 
response time. Thus, we have incrementally indexed batches of 1.000 meta-models (randomly selected), and for each new batch we 
measure the response time for each query mutant.
Moreover, to evaluate the effect of the query size we 
count the number of packages, classes, structural features
and enumerations and we classify
the queries in three types: small (less than 20 elements), medium
(between 20 and 70) and large (more than 70).
Table~\ref{tab:stat} shows some details about the contents
of the queries. 

We run the experiments on a desktop machine with an i7-5820K CPU, with 6 cores at 3.30GHz and maximum heap 16GB. We use Docker to deploy HBase (which runs in a pseudo-distributed mode) and we 
access it locally since we are mainly interested in understanding
the raw performance of our design of the inverted index, and not 
on potential network effects.

\begin{table}[H]
\centering
\footnotesize
\begin{tabular}{c|c|c|c|c|c|c|}
\cline{2-7}
                            & \multicolumn{2}{c|}{Small} & \multicolumn{2}{c|}{Medium} & \multicolumn{2}{c|}{Large} \\ \cline{2-7} 
                            & Mean         & Max         & Mean         & Max          & Mean         & Max         \\ \hline
\multicolumn{1}{|c|}{Paths} & 0            & 0.07        & 0.01         & 0.17         & 0.06         & 0.79        \\ \hline
\multicolumn{1}{|c|}{Get}   & 0.31         & 1.07        & 0.78         & 2.85         & 1.38         & 4.63        \\ \hline
\multicolumn{1}{|c|}{Score} & 0.22         & 0.56        & 0.45         & 0.97         & 0.61         & 1.39        \\ \hline
\multicolumn{1}{|c|}{Total} & 0.53         & 1.49        & 1.24         & 3.99         & 2.06         & 6.27        \\ \hline
\end{tabular}
\caption{Results for an index with 17.000 meta-models.}
\label{tab:response_time}
\end{table}

\begin{figure}[h]
    \centering
    \includegraphics[scale=0.5]{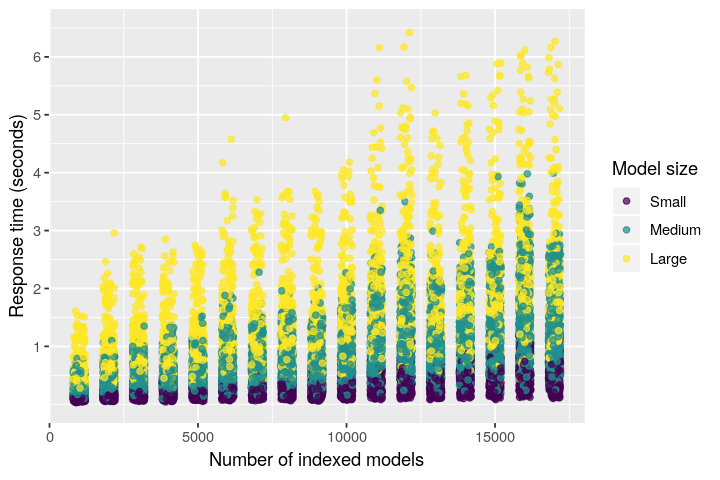}
    \caption{Query response time depending on the index size.}
    \label{fig:execution_time}
\end{figure}

Fig.~\ref{fig:execution_time} shows the relationship between index size and response time, and Table~\ref{tab:response_time} summarizes the results for the largest index.
As can be
observed \mar scales smoothly as the size of the index increases
and, as expected, larger queries take more time (except some outliers because of the garbage collector).
On the other hand, small queries correspond to scenarios in which
the user would develop a query in a fast manner, possibly with
little details. In this case, the average response time is 0.53 seconds for the larger index. The average response time for medium and large queries is 1.24 and 2.06 seconds respectively, which is quite acceptable since there are queries with up to 193 classes.
This would be useful in scenarios like model clone detection.
In general, the computation of the bag of paths and the scoring are 
fast. The execution of {\em get} petitions to HBase consumes most
of the execution time, and it is where we could improve more
the overall performance. Nevertheless, this experiment shows that \mar is already an efficient search engine. 





\section{Related work}\label{sec:related}
In this section we review works related to our proposal,
organised in three categories: model encoding techniques 
which are relevant for model search and indexing, model repositories
and search engines for models themselves.

\begin{table*}[t]
\centering
\footnotesize
\begin{tabular}{l|l|l|l|l|l|l|}
\cline{2-7}
                         & Query format & Search type & Models & Index & Crawler & Megamodel-aware \\ \hline
\multicolumn{1}{|l|}{\cite{lucredio2008moogle} MOOGLE} & Text & Text search             & Any                   & Yes & No &   No           \\ \hline
\multicolumn{1}{|l|}{\cite{barmpis2013hawk} HAWK}   & OCL-like language & Exact results                 & Any                   & Yes & No &  Partial              \\ \hline
\multicolumn{1}{|l|}{\cite{bislimovska2014textual} Text version}   & Text search  & Text search                  & WebML                   & Yes &  No & No            \\ \hline
\multicolumn{1}{|l|}{\cite{bislimovska2014textual} Structure-based version}   & Query-by-example & Structure-based                  & WebML                   & No  & No & No          \\ \hline
\multicolumn{1}{|l|}{\cite{gomes2004using}}   & Query-by-example & Structure-based                  & UML                   & Yes  & No &  No          \\ \hline
\multicolumn{1}{|l|}{\cite{kling2011moscript} MoScript}   & OCL-like language & Exact results               & Any                   & No & No &  Yes            \\ \hline
\multicolumn{1}{|l|}{\cite{basciani2018exploring}}   & Text search & Text search                 & Any                   & Yes  & No & Yes            \\ \hline
\multicolumn{1}{|l|}{\mar}   & Query-by-example & Structure-based               & Any                   & Yes  & No & No            \\ \hline
\end{tabular}
\caption{Summary of search engine for models approaches adapted and extended from \cite{basciani2018exploring}}
\label{tab:engines}
\end{table*}

\vspace{2pt}
{\bf\noindent Model encoding techniques.}
Code2Vec is an technique for representing snippets of code as continuous distributed vectors~\cite{alon2019code2vec}. For each code snipped, a multiset of paths (named \textit{bag of path-contexts}) is extracted from the AST and it is used as an input of a neural network. Our idea of $BoP$ was inspired by the concept of \textit{bag of path-contexts}.
SAMOS is a platform for model analytics, which employs a technique to transform models into vectors in order to apply supervised learning, unsupervised learning, statistical analysis, etc. SAMOS is used in~\cite{babur2017using} for model clustering and meta-model clone detection is proposed in~\cite{babur2019metamodel}. The notion of $n$-grams is used to represent paths of length $n$ between nodes in a graph associated to a model. Models are transformed into a set of $n$-grams and then to a explicit vector whose dimension is determined by the number of distinct $n$-grams in the repository. This is similar to \mar, since we both use paths to encode the structure. However there are some key differences: our graph is different as we consider single attributes as independent nodes, and therefore the extracted paths are different. SAMOS considers paths of the same length, whereas our $BoPs$ are formed by paths of length less or equal than a threshold. 
SAMOS is, in principle, not oriented to implement a search engine since it uses similarity metrics (to build the vectors) which are not thought to perform a fast search.
In~\cite{martinez2018robust} locality sensitive hashing is applied
to summarize EMF models as a hash. This has different applications
including intellectual property protection and fast model comparison. This could be
an alternative strategy to organize a search index. On the other hand, AURORA~\cite{nguyen2019automated} extracts tokens from an Ecore model following an encoding schema (they propose three encoding schemes that try to reflect the model structure) and then, using these tokens, models are transformed to a vector in a vector space that will be an input of a neural network. In \cite{clariso2018applying}, the authors propose using \textit{graph kernels} for clustering software modeling artifacts. The general idea is to transform a set of software models to a set of graphs and then use \textit{graph kernels} to measure the distance between each other.
For clone detection, Exas is a vector encoding of graph models to perform
approximate matching of model fragments~\cite{nguyen2009accurate,pham2009complete}, which has also
been applied to detect model transformation clones~\cite{struber2019model}.
In \cite{kessentini2014search} both structural and syntactic metrics are used in order to compare two meta-models. Also, they used genetic algorithms for meta-model matching. However, their approach did not support indexing so the matching was not though to be fast.
The notion of domain-specific distance is discussed in~\cite{syriani2019domain}, 
which could be incorporated to improve the precision of the search.


\vspace{2pt}
{\bf\noindent Model repositories}.
MDEForge is a collaborative modelling platform~\cite{basciani2014mdeforge} intended to foster ``modelling as a service'' by providing facilities like model storage, search, clustering, workspaces, etc. It is based on a mega-model to keep track of the relationships between the stored modelling artefacts. 
GenMyModel is a cloud service which provides online editors to create different types of models and
stores thousands of public models~\cite{GenMyModel}.
ReMoDD is a repository of MDE artifacts of different nature, which is available through a web-based interface~\cite{france2006repository}.
Hawk provides an architecture to index models~\cite{barmpis2013hawk}, typically for private projects and it can be queried with OCL.
Regarding built-in search facilities in existing model repositories, the study in~\cite{di2015collaborative} shows that search is
typically keyword-based, tag-based or simply there is no search
facility.

\vspace{2pt}
{\bf\noindent Search engines.} Several search engines for models have been proposed in the past
years, but as far as we know none of them is widely used. Table~\ref{tab:engines} shows a summary
of their characteristics.  
In \cite{bislimovska2014textual}, a search engine for WebML models is presented. It supports two query formats: keywords and example-based query. An index is supported only in the keyword query format so structured-based search could be slow if the repository is large. MOOGLE~\cite{lucredio2008moogle} is a generic search engine that only supports text queries in which the user can specify the type of the desired model element to be returned. MOOGLE uses the Apache Lucene query syntax and Apache SOLR as the backend search engine. In \cite{basciani2018exploring}, a similar approach to MOOGLE is proposed with some new features like \textit{megamodel-awareness} (consideration of relations among different kinds of artifacts). On the other hand, \cite{gomes2004using} focuses on the retrieval of UML models using the combination of WordNet and Case-Based Reasoning. 
MoScript is proposed in \cite{kling2011moscript}, a DSL for querying and manipulating model repositories. This approach is model-independent, but the user has to type complex OCL-like queries that only retrieve exact models, not a ranking list of models. 
Regarding our query-by-example approach, some graph transformation engines have been extended to support this feature. For instance, \textsc{IncQuery} has been extended to derive a graph query from an example model~\cite{bergmann2014graph}, and VMQL (Visual Model Query Language) uses the concrete syntax of the source language as the query language~\cite{storrle2011vmql}.



Regarding the scale of the system in terms of the number of models handled, the closest system to \mar is SAMOS \cite{babur2019metamodel}, which performs clustering of 22k meta-models. However, a clustering process is different from indexing, so
the approaches are not really comparable. Nevertheless, clustering a huge model repository is computationally expensive. In particular, SAMOS is running over Apache Spark in \cite{babur2018towards} and takes $17$ hours to process a repository of 7k meta-models. With respect to the size of the indexed models, our system increases in at least an order of magnitude existing proposals. For instance, MOOGLE~\cite{lucredio2008moogle} only indexes 146 models, \cite{bislimovska2014textual} indexes 12 real-world WebML projects which consist on 342 models, and in \cite{basciani2018exploring} up to 3000 models are indexed.







\section{Conclusions}\label{sec:conclusions}
In this work we have presented a novel search engine for models. 
The key idea is to use paths between attribute values to encode
the model structure in an indexable manner. Our evaluation
shows that \mar is both precise and efficient. 
Moreover, we have applied \mar to meta-model classification
obtaining results comparable to state-of-the-art techniques.
Regarding its practical usage,
\mar is able to crawl models in several sources and it has currently indexed more than 50.000 models of different kind, including Ecore meta-models, BPMN diagrams and UML models. This makes it a unique system in the MDE ecosystem, that we hope it is useful for the community.


As future work we plan to improve
our crawlers to reach more repositories. We also
want to include information about the model quality
in the ranking. Finally, as \mar is used in practice we aim at
thoroughly evaluating it with an empirical study.


\section*{Acknowledgments}
Work funded by 
the Spanish Ministry of Science, project \code{TIN2015-73968-JIN AEI/FEDER/UE}
and a Ram\'on y Cajal 2017 grant.

\bibliographystyle{ACM-Reference-Format}
\bibliography{main}


\begin{thebibliography}{39}


\ifx \showCODEN    \undefined \def \showCODEN     #1{\unskip}     \fi
\ifx \showDOI      \undefined \def \showDOI       #1{#1}\fi
\ifx \showISBNx    \undefined \def \showISBNx     #1{\unskip}     \fi
\ifx \showISBNxiii \undefined \def \showISBNxiii  #1{\unskip}     \fi
\ifx \showISSN     \undefined \def \showISSN      #1{\unskip}     \fi
\ifx \showLCCN     \undefined \def \showLCCN      #1{\unskip}     \fi
\ifx \shownote     \undefined \def \shownote      #1{#1}          \fi
\ifx \showarticletitle \undefined \def \showarticletitle #1{#1}   \fi
\ifx \showURL      \undefined \def \showURL       {\relax}        \fi
\providecommand\bibfield[2]{#2}
\providecommand\bibinfo[2]{#2}
\providecommand\natexlab[1]{#1}
\providecommand\showeprint[2][]{arXiv:#2}

\bibitem[\protect\citeauthoryear{??}{Gen}{[n.d.]}]%
        {GenMyModel}
 \bibinfo{year}{[n.d.]}\natexlab{}.
\newblock \bibinfo{title}{GenMyModel}.
\newblock \bibinfo{howpublished}{\url{https://www.genmymodel.com/}}.
\newblock


\bibitem[\protect\citeauthoryear{??}{Who}{[n.d.]}]%
        {Whoosh}
 \bibinfo{year}{[n.d.]}\natexlab{}.
\newblock \bibinfo{title}{Whoosh}.
\newblock
  \bibinfo{howpublished}{\url{https://whoosh.readthedocs.io/en/latest/}}.
\newblock


\bibitem[\protect\citeauthoryear{Alon, Zilberstein, Levy, and Yahav}{Alon
  et~al\mbox{.}}{2019}]%
        {alon2019code2vec}
\bibfield{author}{\bibinfo{person}{Uri Alon}, \bibinfo{person}{Meital
  Zilberstein}, \bibinfo{person}{Omer Levy}, {and} \bibinfo{person}{Eran
  Yahav}.} \bibinfo{year}{2019}\natexlab{}.
\newblock \showarticletitle{code2vec: Learning distributed representations of
  code}.
\newblock \bibinfo{journal}{\emph{Proceedings of the ACM on Programming
  Languages}} \bibinfo{volume}{3}, \bibinfo{number}{POPL}
  (\bibinfo{year}{2019}), \bibinfo{pages}{1--29}.
\newblock


\bibitem[\protect\citeauthoryear{Arasu, Cho, Garcia-Molina, Paepcke, and
  Raghavan}{Arasu et~al\mbox{.}}{2001}]%
        {arasu2001searching}
\bibfield{author}{\bibinfo{person}{Arvind Arasu}, \bibinfo{person}{Junghoo
  Cho}, \bibinfo{person}{Hector Garcia-Molina}, \bibinfo{person}{Andreas
  Paepcke}, {and} \bibinfo{person}{Sriram Raghavan}.}
  \bibinfo{year}{2001}\natexlab{}.
\newblock \showarticletitle{Searching the web}.
\newblock \bibinfo{journal}{\emph{ACM Transactions on Internet Technology
  (TOIT)}} \bibinfo{volume}{1}, \bibinfo{number}{1} (\bibinfo{year}{2001}),
  \bibinfo{pages}{2--43}.
\newblock


\bibitem[\protect\citeauthoryear{Babur}{Babur}{2019}]%
        {onder_babur_2019_2585456}
\bibfield{author}{\bibinfo{person}{\"Onder Babur}.}
  \bibinfo{year}{2019}\natexlab{}.
\newblock \bibinfo{booktitle}{\emph{{A labeled Ecore metamodel dataset for
  domain clustering}}}.
\newblock
\urldef\tempurl%
\url{https://doi.org/10.5281/zenodo.2585456}
\showDOI{\tempurl}


\bibitem[\protect\citeauthoryear{Babur and Cleophas}{Babur and
  Cleophas}{2017}]%
        {babur2017using}
\bibfield{author}{\bibinfo{person}{{\"O}nder Babur} {and} \bibinfo{person}{Loek
  Cleophas}.} \bibinfo{year}{2017}\natexlab{}.
\newblock \showarticletitle{Using n-grams for the Automated Clustering of
  Structural Models}. In \bibinfo{booktitle}{\emph{International Conference on
  Current Trends in Theory and Practice of Informatics}}. Springer,
  \bibinfo{pages}{510--524}.
\newblock


\bibitem[\protect\citeauthoryear{Babur, Cleophas, and van~den Brand}{Babur
  et~al\mbox{.}}{2016}]%
        {babur2016hierarchical}
\bibfield{author}{\bibinfo{person}{{\"O}nder Babur}, \bibinfo{person}{Loek
  Cleophas}, {and} \bibinfo{person}{Mark van~den Brand}.}
  \bibinfo{year}{2016}\natexlab{}.
\newblock \showarticletitle{Hierarchical clustering of metamodels for
  comparative analysis and visualization}. In
  \bibinfo{booktitle}{\emph{European Conference on Modelling Foundations and
  Applications}}. Springer, \bibinfo{pages}{3--18}.
\newblock


\bibitem[\protect\citeauthoryear{Babur, Cleophas, and van~den Brand}{Babur
  et~al\mbox{.}}{2018}]%
        {babur2018towards}
\bibfield{author}{\bibinfo{person}{{\"O}nder Babur}, \bibinfo{person}{Loek
  Cleophas}, {and} \bibinfo{person}{Mark van~den Brand}.}
  \bibinfo{year}{2018}\natexlab{}.
\newblock \showarticletitle{Towards Distributed Model Analytics with Apache
  Spark.}. In \bibinfo{booktitle}{\emph{MODELSWARD}}.
  \bibinfo{pages}{767--772}.
\newblock


\bibitem[\protect\citeauthoryear{Babur, Cleophas, and van~den Brand}{Babur
  et~al\mbox{.}}{2019}]%
        {babur2019metamodel}
\bibfield{author}{\bibinfo{person}{{\"O}nder Babur}, \bibinfo{person}{Loek
  Cleophas}, {and} \bibinfo{person}{Mark van~den Brand}.}
  \bibinfo{year}{2019}\natexlab{}.
\newblock \showarticletitle{Metamodel clone detection with SAMOS}.
\newblock \bibinfo{journal}{\emph{Journal of Computer Languages}}
  (\bibinfo{year}{2019}).
\newblock


\bibitem[\protect\citeauthoryear{Barmpis and Kolovos}{Barmpis and
  Kolovos}{2013}]%
        {barmpis2013hawk}
\bibfield{author}{\bibinfo{person}{Konstantinos Barmpis} {and}
  \bibinfo{person}{Dimitris Kolovos}.} \bibinfo{year}{2013}\natexlab{}.
\newblock \showarticletitle{Hawk: Towards a scalable model indexing
  architecture}. In \bibinfo{booktitle}{\emph{Proceedings of the Workshop on
  Scalability in Model Driven Engineering}}. \bibinfo{pages}{1--9}.
\newblock


\bibitem[\protect\citeauthoryear{Basciani, Di~Rocco, Di~Ruscio, Di~Salle,
  Iovino, and Pierantonio}{Basciani et~al\mbox{.}}{2014}]%
        {basciani2014mdeforge}
\bibfield{author}{\bibinfo{person}{Francesco Basciani}, \bibinfo{person}{Juri
  Di~Rocco}, \bibinfo{person}{Davide Di~Ruscio}, \bibinfo{person}{Amleto
  Di~Salle}, \bibinfo{person}{Ludovico Iovino}, {and} \bibinfo{person}{Alfonso
  Pierantonio}.} \bibinfo{year}{2014}\natexlab{}.
\newblock \showarticletitle{MDEForge: an Extensible Web-Based Modeling
  Platform.}. In \bibinfo{booktitle}{\emph{CloudMDE@ MoDELS}}.
  \bibinfo{pages}{66--75}.
\newblock


\bibitem[\protect\citeauthoryear{Basciani, Di~Rocco, Di~Ruscio, Iovino, and
  Pierantonio}{Basciani et~al\mbox{.}}{2015}]%
        {basciani2015model}
\bibfield{author}{\bibinfo{person}{Francesco Basciani}, \bibinfo{person}{Juri
  Di~Rocco}, \bibinfo{person}{Davide Di~Ruscio}, \bibinfo{person}{Ludovico
  Iovino}, {and} \bibinfo{person}{Alfonso Pierantonio}.}
  \bibinfo{year}{2015}\natexlab{}.
\newblock \showarticletitle{Model repositories: Will they become reality?}. In
  \bibinfo{booktitle}{\emph{CloudMDE@ MoDELS}}. \bibinfo{pages}{37--42}.
\newblock


\bibitem[\protect\citeauthoryear{Basciani, Di~Rocco, Di~Ruscio, Iovino, and
  Pierantonio}{Basciani et~al\mbox{.}}{2016}]%
        {basciani2016automated}
\bibfield{author}{\bibinfo{person}{Francesco Basciani}, \bibinfo{person}{Juri
  Di~Rocco}, \bibinfo{person}{Davide Di~Ruscio}, \bibinfo{person}{Ludovico
  Iovino}, {and} \bibinfo{person}{Alfonso Pierantonio}.}
  \bibinfo{year}{2016}\natexlab{}.
\newblock \showarticletitle{Automated clustering of metamodel repositories}. In
  \bibinfo{booktitle}{\emph{International Conference on Advanced Information
  Systems Engineering}}. Springer, \bibinfo{pages}{342--358}.
\newblock


\bibitem[\protect\citeauthoryear{Basciani, Di~Rocco, Di~Ruscio, Iovino, and
  Pierantonio}{Basciani et~al\mbox{.}}{2018}]%
        {basciani2018exploring}
\bibfield{author}{\bibinfo{person}{Francesco Basciani}, \bibinfo{person}{Juri
  Di~Rocco}, \bibinfo{person}{Davide Di~Ruscio}, \bibinfo{person}{Ludovico
  Iovino}, {and} \bibinfo{person}{Alfonso Pierantonio}.}
  \bibinfo{year}{2018}\natexlab{}.
\newblock \showarticletitle{Exploring model repositories by means of
  megamodel-aware search operators.}. In \bibinfo{booktitle}{\emph{MODELS
  Workshops}}. \bibinfo{pages}{793--798}.
\newblock


\bibitem[\protect\citeauthoryear{Bergmann, Heged{\"u}s, Gerencs{\'e}r, and
  Varr{\'o}}{Bergmann et~al\mbox{.}}{2014}]%
        {bergmann2014graph}
\bibfield{author}{\bibinfo{person}{G{\'a}bor Bergmann},
  \bibinfo{person}{{\'A}bel Heged{\"u}s}, \bibinfo{person}{Gy{\"o}rgy
  Gerencs{\'e}r}, {and} \bibinfo{person}{D{\'a}niel Varr{\'o}}.}
  \bibinfo{year}{2014}\natexlab{}.
\newblock \showarticletitle{Graph Query by Example.}. In
  \bibinfo{booktitle}{\emph{CMSEBA@ MoDELS}}. \bibinfo{pages}{17--24}.
\newblock


\bibitem[\protect\citeauthoryear{Bislimovska, Bozzon, Brambilla, and
  Fraternali}{Bislimovska et~al\mbox{.}}{2014}]%
        {bislimovska2014textual}
\bibfield{author}{\bibinfo{person}{Bojana Bislimovska},
  \bibinfo{person}{Alessandro Bozzon}, \bibinfo{person}{Marco Brambilla}, {and}
  \bibinfo{person}{Piero Fraternali}.} \bibinfo{year}{2014}\natexlab{}.
\newblock \showarticletitle{Textual and content-based search in repositories of
  web application models}.
\newblock \bibinfo{journal}{\emph{ACM Transactions on the Web (TWEB)}}
  \bibinfo{volume}{8}, \bibinfo{number}{2} (\bibinfo{year}{2014}),
  \bibinfo{pages}{1--47}.
\newblock


\bibitem[\protect\citeauthoryear{Bucchiarone, Cabot, Paige, and
  Pierantonio}{Bucchiarone et~al\mbox{.}}{2020}]%
        {bucchiarone2020grand}
\bibfield{author}{\bibinfo{person}{Antonio Bucchiarone}, \bibinfo{person}{Jordi
  Cabot}, \bibinfo{person}{Richard~F Paige}, {and} \bibinfo{person}{Alfonso
  Pierantonio}.} \bibinfo{year}{2020}\natexlab{}.
\newblock \showarticletitle{Grand challenges in model-driven engineering: an
  analysis of the state of the research}.
\newblock \bibinfo{journal}{\emph{Software and Systems Modeling}}
  (\bibinfo{year}{2020}), \bibinfo{pages}{1--9}.
\newblock


\bibitem[\protect\citeauthoryear{Chang, Dean, Ghemawat, Hsieh, Wallach,
  Burrows, Chandra, Fikes, and Gruber}{Chang et~al\mbox{.}}{2008}]%
        {chang2008bigtable}
\bibfield{author}{\bibinfo{person}{Fay Chang}, \bibinfo{person}{Jeffrey Dean},
  \bibinfo{person}{Sanjay Ghemawat}, \bibinfo{person}{Wilson~C Hsieh},
  \bibinfo{person}{Deborah~A Wallach}, \bibinfo{person}{Mike Burrows},
  \bibinfo{person}{Tushar Chandra}, \bibinfo{person}{Andrew Fikes}, {and}
  \bibinfo{person}{Robert~E Gruber}.} \bibinfo{year}{2008}\natexlab{}.
\newblock \showarticletitle{Bigtable: A distributed storage system for
  structured data}.
\newblock \bibinfo{journal}{\emph{ACM Transactions on Computer Systems (TOCS)}}
  \bibinfo{volume}{26}, \bibinfo{number}{2} (\bibinfo{year}{2008}),
  \bibinfo{pages}{1--26}.
\newblock


\bibitem[\protect\citeauthoryear{Claris{\'o} and Cabot}{Claris{\'o} and
  Cabot}{2018}]%
        {clariso2018applying}
\bibfield{author}{\bibinfo{person}{Robert Claris{\'o}} {and}
  \bibinfo{person}{Jordi Cabot}.} \bibinfo{year}{2018}\natexlab{}.
\newblock \showarticletitle{Applying graph kernels to model-driven engineering
  problems}. In \bibinfo{booktitle}{\emph{Proceedings of the 1st International
  Workshop on Machine Learning and Software Engineering in Symbiosis}}.
  \bibinfo{pages}{1--5}.
\newblock


\bibitem[\protect\citeauthoryear{Di~Rocco, Di~Ruscio, Iovino, and
  Pierantonio}{Di~Rocco et~al\mbox{.}}{2015}]%
        {di2015collaborative}
\bibfield{author}{\bibinfo{person}{Juri Di~Rocco}, \bibinfo{person}{Davide
  Di~Ruscio}, \bibinfo{person}{Ludovico Iovino}, {and} \bibinfo{person}{Alfonso
  Pierantonio}.} \bibinfo{year}{2015}\natexlab{}.
\newblock \showarticletitle{Collaborative repositories in model-driven
  engineering [software technology]}.
\newblock \bibinfo{journal}{\emph{IEEE Software}} \bibinfo{volume}{32},
  \bibinfo{number}{3} (\bibinfo{year}{2015}), \bibinfo{pages}{28--34}.
\newblock


\bibitem[\protect\citeauthoryear{France, Bieman, and Cheng}{France
  et~al\mbox{.}}{2006}]%
        {france2006repository}
\bibfield{author}{\bibinfo{person}{Robert France}, \bibinfo{person}{Jim
  Bieman}, {and} \bibinfo{person}{Betty~HC Cheng}.}
  \bibinfo{year}{2006}\natexlab{}.
\newblock \showarticletitle{Repository for model driven development (ReMoDD)}.
  In \bibinfo{booktitle}{\emph{International Conference on Model Driven
  Engineering Languages and Systems}}. Springer, \bibinfo{pages}{311--317}.
\newblock


\bibitem[\protect\citeauthoryear{George}{George}{2011}]%
        {george2011hbase}
\bibfield{author}{\bibinfo{person}{Lars George}.}
  \bibinfo{year}{2011}\natexlab{}.
\newblock \bibinfo{booktitle}{\emph{HBase: the definitive guide: random access
  to your planet-size data}}.
\newblock \bibinfo{publisher}{" O'Reilly Media, Inc."}.
\newblock


\bibitem[\protect\citeauthoryear{Gomes, Pereira, Paiva, Seco, Carreiro,
  Ferreira, and Bento}{Gomes et~al\mbox{.}}{2004}]%
        {gomes2004using}
\bibfield{author}{\bibinfo{person}{Paulo Gomes}, \bibinfo{person}{Francisco~C
  Pereira}, \bibinfo{person}{Paulo Paiva}, \bibinfo{person}{Nuno Seco},
  \bibinfo{person}{Paulo Carreiro}, \bibinfo{person}{Jos{\'e}~L Ferreira},
  {and} \bibinfo{person}{Carlos Bento}.} \bibinfo{year}{2004}\natexlab{}.
\newblock \showarticletitle{Using WordNet for case-based retrieval of UML
  models}.
\newblock \bibinfo{journal}{\emph{AI Communications}} \bibinfo{volume}{17},
  \bibinfo{number}{1} (\bibinfo{year}{2004}), \bibinfo{pages}{13--23}.
\newblock


\bibitem[\protect\citeauthoryear{Kessentini, Ouni, Langer, Wimmer, and
  Bechikh}{Kessentini et~al\mbox{.}}{2014}]%
        {kessentini2014search}
\bibfield{author}{\bibinfo{person}{Marouane Kessentini}, \bibinfo{person}{Ali
  Ouni}, \bibinfo{person}{Philip Langer}, \bibinfo{person}{Manuel Wimmer},
  {and} \bibinfo{person}{Slim Bechikh}.} \bibinfo{year}{2014}\natexlab{}.
\newblock \showarticletitle{Search-based metamodel matching with structural and
  syntactic measures}.
\newblock \bibinfo{journal}{\emph{Journal of Systems and Software}}
  \bibinfo{volume}{97} (\bibinfo{year}{2014}), \bibinfo{pages}{1--14}.
\newblock


\bibitem[\protect\citeauthoryear{Kling, Jouault, Wagelaar, Brambilla, and
  Cabot}{Kling et~al\mbox{.}}{2011}]%
        {kling2011moscript}
\bibfield{author}{\bibinfo{person}{Wolfgang Kling},
  \bibinfo{person}{Fr{\'e}d{\'e}ric Jouault}, \bibinfo{person}{Dennis
  Wagelaar}, \bibinfo{person}{Marco Brambilla}, {and} \bibinfo{person}{Jordi
  Cabot}.} \bibinfo{year}{2011}\natexlab{}.
\newblock \showarticletitle{MoScript: A DSL for querying and manipulating model
  repositories}. In \bibinfo{booktitle}{\emph{International Conference on
  Software Language Engineering}}. Springer, \bibinfo{pages}{180--200}.
\newblock


\bibitem[\protect\citeauthoryear{Lucr{\'e}dio, Fortes, and
  Whittle}{Lucr{\'e}dio et~al\mbox{.}}{2008}]%
        {lucredio2008moogle}
\bibfield{author}{\bibinfo{person}{Daniel Lucr{\'e}dio},
  \bibinfo{person}{Renata P de~M Fortes}, {and} \bibinfo{person}{Jon Whittle}.}
  \bibinfo{year}{2008}\natexlab{}.
\newblock \showarticletitle{MOOGLE: A model search engine}. In
  \bibinfo{booktitle}{\emph{International Conference on Model Driven
  Engineering Languages and Systems}}. Springer, \bibinfo{pages}{296--310}.
\newblock


\bibitem[\protect\citeauthoryear{Lucr{\'e}dio, Fortes, and
  Whittle}{Lucr{\'e}dio et~al\mbox{.}}{2012}]%
        {lucredio2012moogle}
\bibfield{author}{\bibinfo{person}{Daniel Lucr{\'e}dio},
  \bibinfo{person}{Renata P de~M Fortes}, {and} \bibinfo{person}{Jon Whittle}.}
  \bibinfo{year}{2012}\natexlab{}.
\newblock \showarticletitle{{MOOGLE}: a metamodel-based model search engine}.
\newblock \bibinfo{journal}{\emph{Software \& Systems Modeling}}
  \bibinfo{volume}{11}, \bibinfo{number}{2} (\bibinfo{year}{2012}),
  \bibinfo{pages}{183--208}.
\newblock


\bibitem[\protect\citeauthoryear{Mart{\'\i}nez, G{\'e}rard, and
  Cabot}{Mart{\'\i}nez et~al\mbox{.}}{2018}]%
        {martinez2018robust}
\bibfield{author}{\bibinfo{person}{Salvador Mart{\'\i}nez},
  \bibinfo{person}{S{\'e}bastien G{\'e}rard}, {and} \bibinfo{person}{Jordi
  Cabot}.} \bibinfo{year}{2018}\natexlab{}.
\newblock \showarticletitle{Robust hashing for models}. In
  \bibinfo{booktitle}{\emph{Proceedings of the 21th ACM/IEEE International
  Conference on Model Driven Engineering Languages and Systems}}.
  \bibinfo{pages}{312--322}.
\newblock


\bibitem[\protect\citeauthoryear{Nguyen, Nguyen, Pham, Al-Kofahi, and
  Nguyen}{Nguyen et~al\mbox{.}}{2009}]%
        {nguyen2009accurate}
\bibfield{author}{\bibinfo{person}{Hoan~Anh Nguyen},
  \bibinfo{person}{Tung~Thanh Nguyen}, \bibinfo{person}{Nam~H Pham},
  \bibinfo{person}{Jafar~M Al-Kofahi}, {and} \bibinfo{person}{Tien~N Nguyen}.}
  \bibinfo{year}{2009}\natexlab{}.
\newblock \showarticletitle{Accurate and efficient structural characteristic
  feature extraction for clone detection}. In
  \bibinfo{booktitle}{\emph{International Conference on Fundamental Approaches
  to Software Engineering}}. Springer, \bibinfo{pages}{440--455}.
\newblock


\bibitem[\protect\citeauthoryear{Nguyen, Di~Rocco, Di~Ruscio, Pierantonio, and
  Iovino}{Nguyen et~al\mbox{.}}{2019}]%
        {nguyen2019automated}
\bibfield{author}{\bibinfo{person}{Phuong~T Nguyen}, \bibinfo{person}{Juri
  Di~Rocco}, \bibinfo{person}{Davide Di~Ruscio}, \bibinfo{person}{Alfonso
  Pierantonio}, {and} \bibinfo{person}{Ludovico Iovino}.}
  \bibinfo{year}{2019}\natexlab{}.
\newblock \showarticletitle{Automated Classification of Metamodel Repositories:
  A Machine Learning Approach}. In \bibinfo{booktitle}{\emph{2019 ACM/IEEE 22nd
  International Conference on Model Driven Engineering Languages and Systems
  (MODELS)}}. IEEE, \bibinfo{pages}{272--282}.
\newblock


\bibitem[\protect\citeauthoryear{Pham, Nguyen, Nguyen, Al-Kofahi, and
  Nguyen}{Pham et~al\mbox{.}}{2009}]%
        {pham2009complete}
\bibfield{author}{\bibinfo{person}{Nam~H Pham}, \bibinfo{person}{Hoan~Anh
  Nguyen}, \bibinfo{person}{Tung~Thanh Nguyen}, \bibinfo{person}{Jafar~M
  Al-Kofahi}, {and} \bibinfo{person}{Tien~N Nguyen}.}
  \bibinfo{year}{2009}\natexlab{}.
\newblock \showarticletitle{Complete and accurate clone detection in
  graph-based models}. In \bibinfo{booktitle}{\emph{2009 IEEE 31st
  International Conference on Software Engineering}}. IEEE,
  \bibinfo{pages}{276--286}.
\newblock


\bibitem[\protect\citeauthoryear{Porter}{Porter}{1980}]%
        {Porter1980AnAF}
\bibfield{author}{\bibinfo{person}{Martin~F. Porter}.}
  \bibinfo{year}{1980}\natexlab{}.
\newblock \showarticletitle{An algorithm for suffix stripping}.
\newblock \bibinfo{journal}{\emph{Program}}  \bibinfo{volume}{40}
  (\bibinfo{year}{1980}), \bibinfo{pages}{211--218}.
\newblock


\bibitem[\protect\citeauthoryear{Robertson, Zaragoza, et~al\mbox{.}}{Robertson
  et~al\mbox{.}}{2009}]%
        {robertson2009probabilistic}
\bibfield{author}{\bibinfo{person}{Stephen Robertson}, \bibinfo{person}{Hugo
  Zaragoza}, {et~al\mbox{.}}} \bibinfo{year}{2009}\natexlab{}.
\newblock \showarticletitle{The probabilistic relevance framework: BM25 and
  beyond}.
\newblock \bibinfo{journal}{\emph{Foundations and Trends{\textregistered} in
  Information Retrieval}} \bibinfo{volume}{3}, \bibinfo{number}{4}
  (\bibinfo{year}{2009}), \bibinfo{pages}{333--389}.
\newblock


\bibitem[\protect\citeauthoryear{Robles, Ho-Quang, Hebig, Chaudron, and
  Fernandez}{Robles et~al\mbox{.}}{2017}]%
        {robles2017extensive}
\bibfield{author}{\bibinfo{person}{Gregorio Robles}, \bibinfo{person}{Truong
  Ho-Quang}, \bibinfo{person}{Regina Hebig}, \bibinfo{person}{Michel~RV
  Chaudron}, {and} \bibinfo{person}{Miguel~Angel Fernandez}.}
  \bibinfo{year}{2017}\natexlab{}.
\newblock \showarticletitle{An extensive dataset of UML models in GitHub}. In
  \bibinfo{booktitle}{\emph{2017 IEEE/ACM 14th International Conference on
  Mining Software Repositories (MSR)}}. IEEE, \bibinfo{pages}{519--522}.
\newblock


\bibitem[\protect\citeauthoryear{Steinberg, Budinsky, Merks, and
  Paternostro}{Steinberg et~al\mbox{.}}{2008}]%
        {steinberg2008emf}
\bibfield{author}{\bibinfo{person}{Dave Steinberg}, \bibinfo{person}{Frank
  Budinsky}, \bibinfo{person}{Ed Merks}, {and} \bibinfo{person}{Marcelo
  Paternostro}.} \bibinfo{year}{2008}\natexlab{}.
\newblock \bibinfo{booktitle}{\emph{EMF: eclipse modeling framework}}.
\newblock \bibinfo{publisher}{Pearson Education}.
\newblock


\bibitem[\protect\citeauthoryear{St{\"o}rrle}{St{\"o}rrle}{2011}]%
        {storrle2011vmql}
\bibfield{author}{\bibinfo{person}{Harald St{\"o}rrle}.}
  \bibinfo{year}{2011}\natexlab{}.
\newblock \showarticletitle{VMQL: A visual language for ad-hoc model querying}.
\newblock \bibinfo{journal}{\emph{Journal of Visual Languages \& Computing}}
  \bibinfo{volume}{22}, \bibinfo{number}{1} (\bibinfo{year}{2011}),
  \bibinfo{pages}{3--29}.
\newblock


\bibitem[\protect\citeauthoryear{Str{\"u}ber, Acre{\c{t}}oaie, and
  Pl{\"o}ger}{Str{\"u}ber et~al\mbox{.}}{2019}]%
        {struber2019model}
\bibfield{author}{\bibinfo{person}{Daniel Str{\"u}ber}, \bibinfo{person}{Vlad
  Acre{\c{t}}oaie}, {and} \bibinfo{person}{Jennifer Pl{\"o}ger}.}
  \bibinfo{year}{2019}\natexlab{}.
\newblock \showarticletitle{Model clone detection for rule-based model
  transformation languages}.
\newblock \bibinfo{journal}{\emph{Software \& Systems Modeling}}
  \bibinfo{volume}{18}, \bibinfo{number}{2} (\bibinfo{year}{2019}),
  \bibinfo{pages}{995--1016}.
\newblock


\bibitem[\protect\citeauthoryear{Syriani, Bill, and Wimmer}{Syriani
  et~al\mbox{.}}{2019}]%
        {syriani2019domain}
\bibfield{author}{\bibinfo{person}{Eugene Syriani}, \bibinfo{person}{Robert
  Bill}, {and} \bibinfo{person}{Manuel Wimmer}.}
  \bibinfo{year}{2019}\natexlab{}.
\newblock \showarticletitle{Domain-Specific Model Distance Measures}.
\newblock \bibinfo{journal}{\emph{Journal of Object Technology}}
  \bibinfo{volume}{18}, \bibinfo{number}{3} (\bibinfo{year}{2019}).
\newblock


\bibitem[\protect\citeauthoryear{Zhai and Massung}{Zhai and Massung}{2016}]%
        {Zhai2016TextDM}
\bibfield{author}{\bibinfo{person}{ChengXiang Zhai} {and} \bibinfo{person}{Sean
  Massung}.} \bibinfo{year}{2016}\natexlab{}.
\newblock \showarticletitle{Text Data Management and Analysis: A Practical
  Introduction to Information Retrieval and Text Mining}.
\newblock


\end{thebibliography}
\end{document}